\documentclass[]{aa}
\usepackage{hyperref}
\usepackage{txfonts}
\usepackage{caption}
\usepackage{amssymb,tabularx}
\usepackage{amsmath}
\usepackage{graphicx,epsfig,color,latexsym}
\usepackage{pifont}
\usepackage{subcaption}
\usepackage{array}
\renewcommand{\exp}{\mathrm{exp}}

\newcommand{\bea}{\begin{eqnarray}}
\newcommand{\eea}{\end{eqnarray}}
\newcommand{\be}{\begin{equation}}
\newcommand{\ee}{\end{equation}}
\newcommand{\rund}[1]{\left(#1\right)}

\begin{document}

\title{Plasma lensing in comparison to gravitational lensing \\ Formalism and degeneracies}
\titlerunning{Plasma lensing in comparison to gravitational lensing}
\author{Jenny Wagner\inst{1}  \and Xinzhong Er\inst{2}}
\institute{Universit\"at Heidelberg, Zentrum f\"ur Astronomie, Astron.~Rechen-Institut, M\"onchhofstr. 12--14, 69120 Heidelberg, Germany\\
\email{j.wagner@uni-heidelberg.de}
\and
South-Western Institute for Astronomy Research, Yunnan University, 650091, Kunming, Yunnan, P.~R.~China \\
\email{xer@ynu.edu.cn}
}
\date{Received XX; accepted XX}

\abstract{Gravitational and plasma lensing share the same mathematical formalism in the limit of geometrical optics. 
We show that both phenomena can be effectively described by a projected, two-dimensional deflection potential whose gradient causes an instantaneous light deflection in a single, thin lens plane. 
Subsequently, we highlight the differences in the time-delay and lensing equations that occur because plasma lensing is caused by a potential directly proportional to the deflecting electron number density and gravitational lensing is caused by a potential that is related to the deflecting mass density by a Poisson equation. 
Since we treat plasma and gravitational lensing as thin-screen effective theories, their degeneracies are both caused by the unknown distribution of deflecting objects.
Hence, deriving the formalism-intrinsic degeneracies for plasma lensing, we find that they are analogous to those occurring in gravitational lensing.
To break these degeneracies, galaxies and galaxy-cluster scale strong gravitational lenses must rely on additional assumptions or complementary observations. Physically realistic assumptions to arrive at self-consistent lens and source reconstructions can be provided by simulations and analytical effective theories.
In plasma lensing, a deeper understanding of the deflecting electron density distributions is still under development, so that a model-based comprehensive lens reconstruction is not yet possible. However, we show that transient nearby lenses and multi-wavelength observations help to break the arising degeneracies.
We thus conclude that the development of an observation-based inference of local lens properties seems currently the best way to further probe the morphologies of plasma electron densities. 
Furthermore, due to the simpler evidence-based breaking of the lensing degeneracies, we expect to obtain tighter constraints on the local plasma electron densities than on the gravitationally deflecting masses.
}
\maketitle

\section{Introduction}

When electro-magnetic signals travel from their origin to the observer, they can be deflected by refraction or diffraction on their way.
The deviations from their unperturbed path depend on the wavelength at emission, the nature of the intervening objects, and the relative distances of the source, these objects, and the observer.

If an intervening object is an inhomogeneous mass density on top of a background mass density, this inhomogeneity disturbs the homogeneous mass density of the cosmic background spacetime.
Consequently, the path of the electro-magnetic signals deviates from their unperturbed path in the background spacetime.
The mass inhomogeneity distorts the travelling light bundle independent of the emitted wavelength and magnifies or demagnifies it.
Thus, the mass density, which is usually projected into a so-called lens plane orthogonal to the line of sight, acts as a gravitational lens.
Observable distortions of the propagating light bundles occur, if the projected, inhomogeneous deflecting mass density is of the order of the critical surface mass density in the lens plane
\begin{equation}
\Sigma_\mathrm{c} = \dfrac{c^2}{4\pi G}\dfrac{D_\mathrm{s}}{D_\mathrm{d} D_\mathrm{ds}}\;, 
\label{eq:sigma_c}
\end{equation}
in which $c$ is the speed of light, $G$ the gravitational constant, $D_\mathrm{s}$ is the (angular diameter) distance from the observer to the source, $D_\mathrm{d}$ is the (angular diameter) distance from the observer to the lens, and $D_\mathrm{ds}$ is the (angular diameter) distance between the lens and the source.
Here, we focus on the strong gravitational lensing effects, i.e.~those cases in which the perturbing mass density generates multiple images of a single background source. Since the discovery of the first pair of multiple images of a background quasar, \cite{bib:Walsh}, strong gravitational lensing has been developed as a robust tool to constrain the total deflecting masses (e.g.~\cite{bib:Bolton}, \cite{bib:Lagattuta} on galaxy scale and e.g.~ \cite{bib:Chiu}, \cite{bib:Coe}, \cite{bib:Lotz} on galaxy-cluster scale), to infer parameters of the cosmological standard model (e.g.~\cite{bib:Collett}, \cite{bib:Dobke}, \cite{bib:Linder}), in particular the Hubble-Lemaître constant, $H_0$ (e.g.~\cite{bib:Chen}, \cite{bib:Refsdal}, \cite{bib:Suyu}, \cite{bib:Wong}), and to probe the nature of dark matter on sub-galactic scales (e.g.~\cite{bib:Despali}, \cite{bib:Gilman}, \cite{bib:Robertson}). 

Observations of multiple images are sparse for galaxy-cluster-scale lenses and show a high degree of symmetry for galaxy-scale lenses, so that mass density reconstructions must rely on additional model assumptions to infer quantities like the total mass of a lens within a given aperture or to determine the slope of the mass density profile at a specific position. 
Due to a multitude of simulations (see \cite{bib:Kuhlen} for an overview), the morphology of matter agglomerations can be represented by heuristically inferred mass density profiles which complement the sparse observational evidence and which allow us to reconstruct the deflecting mass density distribution and the background source object in a self-consistent way. 
Analytical explanations have also been established to support the simulation results, see e.g.~\cite{bib:Nolting}, \cite{bib:Wagner_halo}, so that a useful degree of understanding of gravitationally deflecting mass agglomerations has been reached. 

If the intervening object is an inhomogeneous electron density distribution, forming a cloud of cold plasma, the travelling light bundle is distorted due to the interaction between the plasma and the traversing electro-magnetic signal.
Contrary to the gravitational lensing effect, this effect of plasma lensing depends on the emitted frequency and becomes negligible for frequencies that are not of the order of the plasma frequency
\begin{equation}
\omega_p \equiv \sqrt{\dfrac{e^2 n_\mathrm{e}}{\epsilon_0 m_\mathrm{e}}} = \sqrt{4\pi r_\mathrm{e} c^2 n_\mathrm{e}}\;,
\label{eq:wp}
\end{equation}
in which $e$ is the electron charge, $m_\mathrm{e}$ is the mass of an electron, $r_\mathrm{e}$ is the classical electron radius, $\epsilon_0$ is the vaccum permittivity and $n_e$ is the number density of electrons in the plasma in its unperturbed equilibrium state.
\cite{bib:Fiedler} gave the first interpretation of large fluctuations in radio observations, dubbed ``extreme scattering events'' (ESE), as refractive focussing by discrete plasma structures in the interstellar medium (ISM) of our Galaxy.
Since then, lensing by plasma inhomogeneities has been subject to many theoretical studies which have not only been restricted to plasma structures in the ISM of our own galaxy (e.g. \cite{bib:Bisnovatyi2}, \cite{bib:Bisnovatyi3}, \cite{bib:Clegg}, \cite{bib:Cordes}, \cite{bib:Er2}, \cite{bib:Er3}, \cite{bib:Grillo}, \cite{bib:Gwinn}, \cite{bib:Pen}, \cite{bib:Rogers}, \cite{bib:Simard1}, \cite{bib:Stinebring}). 
Yet, a general understanding and characterisation of plasma structures is lacking and it remains a challenging task to set up an encompassing characterisation for all observations by specific plasma density models. 
Therefore, results are given on a case-by-case basis (apart from the references mentioned above, further examples are found in~\cite{bib:Bannister}, \cite{bib:Cognard}, \cite{bib:Pushkarev}, \cite{bib:Tuntsov}). 

Developing different models and selecting the ones consistent with observations, viable theories for plasma generation and evolution can be corroborated or refuted. 
But, as previous works already found, the observational evidence may not favour a specific plasma density profile, e.g. \cite{bib:Bannister}, so that many models are equally well suited fits to the data.
The same is true for mass density profiles of gravitational lenses, which fit observations equally well, but may be inconsistent with each other in regions without multiple-image constraints where the model assumptions shape the mass density profile \cite{bib:Raney}.
Furthermore, the assumption of highly symmetric electron density profiles may lead to unrealistic, biased conclusions, as has been recently discovered in the case of gravitational lensing with too symmetric mass density profiles (see e.g.~\cite{bib:Gilman}, \cite{bib:Meneghetti1}, \cite{bib:Wagner_quasar}, \cite{bib:Walls}).

To avoid the mutual inconsistencies of lens models, we have been establishing an observation-based approach to reduce the gravitational lens reconstruction to those \emph{local} lens properties that all lens model agree upon.  
It requires only a minimum amount of theoretical model assumptions and it has already been successfully applied to a galaxy- and a galaxy-cluster-scale gravitational lens, see \cite{bib:Wagner_sum} and references therein.

As plasma lenses can be described by the same mathematical formalism, transferring our approach to this application can help to develop a better understanding of the deflecting plasma structures without the need to assume a specific profile for the scattering electron density distribution.
With an increasing amount of available observations, the sparse, local information about the scattering plasma structures can be connected and statistically combined to further develop a theoretical characterisation of scattering electron density distributions.

In this work, we therefore set the foundations to transfer our observation-based gravitational lens reconstruction to plasma lensing by investigating the similarities and differences in the mathematical formalism and in particular in the degeneracies that need to be broken to obtain characteristic information about the deflecting structures.

This work is outlined as follows: 
In Section~\ref{sec:formalism}, we introduce the general single-lens plane light deflection formalism based on geometrical optics that gravitational and plasma lensing have in common. 
In Section~\ref{sec:differences}, we customise the formalism of Section~\ref{sec:formalism} to plasma lensing and briefly mention the gravitational lensing analogue, which is further detailed in, e.g.\ \cite{bib:SEF} or \cite{bib:Petters}. 
Subsequently, we transfer the formalism-intrinsic degeneracies know from the gravitational lensing formalism to plasma lensing in Section~\ref{sec:degeneracies} and give them a physical interpretation.  
Section~\ref{sec:multi_wavelengths} deals with ways to break the plasma lensing degeneracies by using observations of the source for transient plasma lenses and by using observations at multiple wavelengths. 
We apply the findings of Section~\ref{sec:multi_wavelengths} to a simulated Gaussian plasma lens in Section~\ref{sec:example} to investigate the orders of magnitude of the required measurement precision and analyse the resulting accuracy of our results.
Section~\ref{sec:conclusion} summarises all results and gives an overview of future research directions that observations of plasma lensing effects allow us to pursue.

\section{Geometrical optics formalism for light deflection}
\label{sec:formalism}

We now summarise the formalism to describe light deflection by an arbitrarily shaped, deflecting structure, when the deflecting structure can be characterised by a deflection potential and the light propagation can be described by geometrical optics.
Section~\ref{sec:prerequisites} lists the requirements and assumptions that will be used to set up the formulae which characterise the light deflection in Section~\ref{sec:formulae}.
While similar derivations have been pursued for gravitational and plasma lensing separately, we unify their descriptions here.

\subsection{Prerequisites}
\label{sec:prerequisites}

The standard formalism for light deflection that we employ in this work is based on two assumptions.
They simplify the calculations and allow us to extract variables from the theory that can be directly connected to observable data.
As a full general relativistic treatment of lensing does not change the fundamental statements of this section, we refer to \cite{bib:Perlick}, \cite{bib:SEF} and \cite{bib:Synge} for details about the most general equations to describe geometrical optics in curved spacetimes.

The first assumption of our approach states that the light deflection occurs instantaneously at a quasi-stationary, geometrically thin, deflecting structure that is considered as a small perturbation on top of the background density which is embedded in a general Friedmann universe.
The quasi-stationarity implies that the lens is not moving relative to the observer and the source at the time of observation. 
In particular, the lens does not rotate, such that we can characterise the light deflection by introducing a deflection potential. 
Deflecting structures could be extended along the line of sight.
Therefore, if the deflecting structure is not geometrically thin, projecting all deflectors into one lens plane yields an \emph{effective} description of the light deflection by means of a projected deflection potential.
In most cases, observables are integrated quantities along the entire line of sight from the source to the observer, so that this effective description is a valid approximation for a multitude of applications. 
In order to separate the cosmology-dependent part of the potential from the actual, cosmology-independent light deflection, we furthermore scale the projected deflection potential to a dimensionless quantity that describes the light deflection.

Thus, we introduce the projected, dimensionless deflection potential, $\psi(\boldsymbol{x})$ in the lens plane that depends on the angular extensions $\boldsymbol{x} \in \mathbb{R}^2$ over which the deflecting structure extends on the sky. 
Assuming a lens plane instead of a spherical segment of the celestial sphere, implies that the extension of the deflecting structure on the sky is small compared to its distance from the observer.
Being a deflection potential from which further characteristics of the deflector should be inferred, it is reasonable to assume that $\psi(\boldsymbol{x})$ belongs to the class of differentiable functions. 
Usually, taking $\psi(\boldsymbol{x})$ from the class of smooth functions is a sufficient and convenient assumption for most physical applications, so that we assume it in this work. 
A mathematically thorough treatment with weaker assumptions can be found in \cite{bib:Wagner4}. 

The second assumption of the approach is that light propagation obeys the laws of geometrical optics, i.e.~all wavelengths are small compared to the sizes of the deflecting objects.
According to Fermat's principle, light is modelled as rays that move along paths of stationary travel time in this short-wavelength approximation that neglects all wave phenomena of light.


\subsection{Derivation of the light deflection formulae}
\label{sec:formulae}

From the assumptions stated in Section~\ref{sec:prerequisites}, we now derive the formulae that allow us to determine local properties of the deflecting structure by observables. 
Equations~\eqref{eq:lens_equation2}, \eqref{eq:A}, \eqref{eq:time_delay0}, and \eqref{eq:time_delay1} can be employed in cases when the source can be observed without the deflecting structure (i.e.~for a moving deflector).
For deflectors that move too slowly so that the source cannot be observed, we have to eliminate the source positions from these equations and arrive at Equations~\eqref{eq:lens_equation3}, \eqref{eq:time_delay2}, and \eqref{eq:time_delay3} and at a system of equations of Equation~\eqref{eq:lens_equation2} equating the right-hand sides of this equation for different multiple images of the same source.
Setting up the latter is detailed in \cite{bib:Wagner_sum} and references therein. 

With a continuously differentiable deflection potential $\psi(\boldsymbol{x})$, we define the deflection angle in the lens plane
\begin{equation}
\boldsymbol{\alpha}(\boldsymbol{x})  = \nabla \psi(\boldsymbol{x}) \;.
\label{eq:alpha}
\end{equation}
From this relation, it is obvious that $\psi(\boldsymbol{x})$ is a dimensionless quantity.
It is thus derived from a standard three-dimensional potential by a projection into the lens plane and a subsequent scaling.

At the same time, we connect the angular source position $\boldsymbol{y} \in \mathbb{R}^ 2$ in the source plane at distance $D_\mathrm{s}$ from the observer with the observed angular position $\boldsymbol{x} \in \mathbb{R}^2$ of a deflected image in the lens plane at distance $D_\mathrm{d}$ from the observer by stating that
\begin{equation}
\boldsymbol{y} = \boldsymbol{x} - \boldsymbol{\alpha}(\boldsymbol{x}) \;.
\label{eq:lens_equation1}
\end{equation}
Inserting Equation~\eqref{eq:alpha} into Equation~\eqref{eq:lens_equation1}, we arrive at the lens equation
\begin{equation}
\boldsymbol{y} = \boldsymbol{x} - \nabla \psi(\boldsymbol{x}) \;.
\label{eq:lens_equation2}
\end{equation}

Deriving Equation~\eqref{eq:lens_equation2} with respect to $\boldsymbol{x}$, we obtain the distortion matrix at a position $\boldsymbol{x}$ in the lens plane as
\begin{equation}
A(\boldsymbol{x}) = \left( \begin{matrix} 1 -  \psi_{11}(\boldsymbol{x}) & - \psi_{12}(\boldsymbol{x}) \\ -\psi_{12}(\boldsymbol{x}) & 1 - \psi_{22}(\boldsymbol{x}) \end{matrix} \right)  \;, \quad \psi_{ij} \equiv \dfrac{\partial^2 \psi(\boldsymbol{x})}{\partial x_i \partial x_j} \;, \; i,j=1,2 \;,
\end{equation}
which describes the distortions of a light bundle in the vicinity of $\boldsymbol{x}$.
The distortion matrix can be constrained in a vicinity $\mathcal{V}$ of a point $\boldsymbol{x}_0$ where $A(\boldsymbol{x})$ changes slowly, such that it can be approximated as being constant in $\mathcal{V}(\boldsymbol{x}_0)$. 
In that case, vectors in the lens plane that are contained in $\mathcal{V}(\boldsymbol{x}_0)$ relate to vectors in the source plane that are contained in the back-projection of $\mathcal{V}(\boldsymbol{x}_0)$ in the vicinity of $\boldsymbol{y}_0 = \boldsymbol{x}_0 -\boldsymbol{\alpha}(\boldsymbol{x}_0)$ by
\begin{equation}
\boldsymbol{y} - \boldsymbol{y}_0 = A(\boldsymbol{x}_0) \left(\boldsymbol{x} - \boldsymbol{x}_0 \right) \;.
\label{eq:A}
\end{equation} 

The arrival time difference $t$ between a perturbed light ray with travel time $t_\mathrm{p}$ and an unperturbed light ray with travel time $t_\mathrm{u}$ in a Friedmann cosmological background universe can be explained by two parts, such that 
\begin{equation}
t = t_\mathrm{p} - t_\mathrm{u} \equiv t_\psi + t_\mathrm{g} \;.
\label{eq:t}
\end{equation}
The first part of the right-hand side of Equation~\eqref{eq:t}, $t_\psi$, accounts for the time spent travelling through the deflecting structure and being exposed to its deflection potential.
According to Section~\ref{sec:prerequisites}, we can describe the light path through the deflecting structure by a linear potential theory in a Newtonian spacetime. 
We assume that the deflecting structure can be modelled as a refractive medium with a spatially varying refractive index, $n(\boldsymbol{r})$, with $\boldsymbol{r} \in \mathbb{R}^3$.
Since we assume a weak deflecting potential according to the prerequisites of Section~\ref{sec:prerequisites}, the effective refractive index can be expressed as a perturbation around 1, such that the refractive index for a three-dimensional deflection potential characterising the deflecting medium is given by
\begin{equation}
n(\boldsymbol{r}) = 1 - \tilde{\phi}(\boldsymbol{r}) \;.
\label{eq:n}
\end{equation}
The tilde expresses that the potential is scaled to a dimensionless quantity and includes additional prefactors compared to the standard potentials, e.g. like the standard gravitational potential.
Projecting all deflecting objects into the lens plane at angular diameter distance $D_\mathrm{d}$, the effective potential in this lens plane implies an effective refractive index to characterise the medium, which we set up as
\begin{equation}
n_\mathrm{eff}(\boldsymbol{r}) = n_\mathrm{eff}(\boldsymbol{x},D) = 1 - \tilde{\psi}(\boldsymbol{x}) \delta(D-D_\mathrm{d})\;.
\label{eq:neff_psi}
\end{equation} 
To determine the scaling by which $\psi(\boldsymbol{x})$ is obtained as a dimensionless quantity, we introduce the deflection potential $\tilde{\psi}(\boldsymbol{x})$ in Equation~\eqref{eq:neff_psi}, which we will relate to $\psi(\boldsymbol{x})$ of Equation~\eqref{eq:alpha} at the end of this derivation.
As the refractive index scales the phase velocity with respect to the vacuum propagation velocity, a negative potential causes the light propagation to be retarded, while a positive potential advances the phase velocity.
Equations~\eqref{eq:n} with \eqref{eq:neff_psi} can be related to each other by noting that $\tilde{\psi}(\boldsymbol{x})$ can be obtained by integrating $\tilde{\phi}(\boldsymbol{r})$ along the line of sight from the source at proper distance $|\boldsymbol{d}_\mathrm{OS}|$ to the observer $O$ and converting the remaining distance coordinates in the lens plane to angular quantities. We denote the proper distance over which we integrate by $l$, such that we obtain
\begin{equation}
\tilde{\psi}(\boldsymbol{x}) = \int \limits_{|\boldsymbol{d}_\mathrm{OS}|}^{0} \mathrm{d} l \, \tilde{\phi}(\boldsymbol{r}) \;.
\label{eq:psi_phi}
\end{equation}
The time difference between a light ray travelling through the deflecting medium compared to a light ray travelling through vacuum is given by\footnote{In order to be able to use the $\delta$-distribution in $n_\mathrm{eff}(\boldsymbol{r})$, we extended the limits of the integral to the entire line of sight, instead of inserting the source and observer position. 
Yet, the medium is assumed to be of finite extent close to $D_\mathrm{d}$, such that the mathematically necessary change in the boundary conditions does not have any physically relevant consequences.}
\begin{equation}
t_\psi = \dfrac{(1+z_\mathrm{d})}{c} \int \limits_{-\infty}^{+\infty} \mathrm{d} D \left( n_\mathrm{eff}(\boldsymbol{r}) -1 \right) \;,
\end{equation}
in which the factor of $(1+z_\mathrm{d})$ occurs when transforming the comoving coordinate $l$ to an angular diameter coordinate $D$.
Inserting Equation~\eqref{eq:neff_psi} yields
\begin{equation}
t_\psi = -\dfrac{(1+z_\mathrm{d})}{c} \tilde{\psi}(\boldsymbol{x}) \;.
\label{eq:t_psi}
\end{equation}

\begin{figure}[t!]
\centering
  \includegraphics[width=0.36\textwidth]{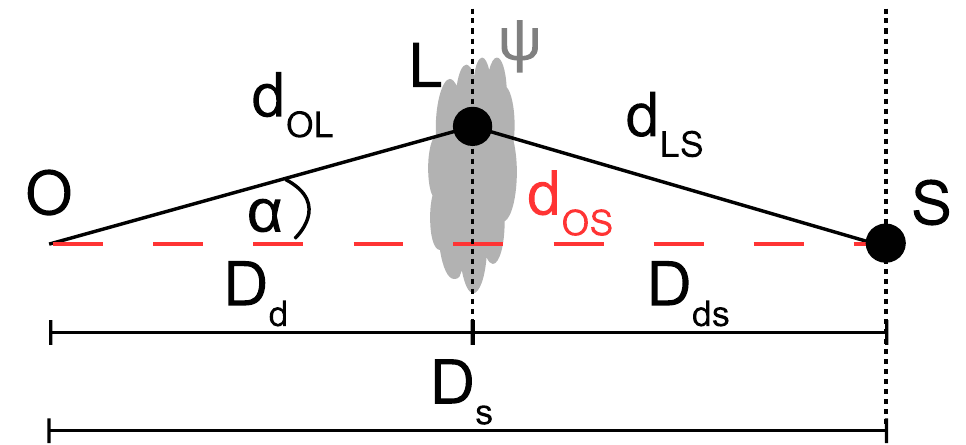} 
    \caption{In the presence of a deflecting structure in the lens plane, represented by a projected deflection potential $\psi$, a light ray emitted from the source $S$ at angular position $\boldsymbol{y}$ in the source plane passes the deflected image of the source $L$ at angular position $\boldsymbol{x}$ in the lens plane before it arrives at the observer $O$. The travel time is assumed to be $t_\mathrm{p}$. Without the deflecting structure at $D_\mathrm{d}$, the light ray travels the angular diameter distance $D_\mathrm{s}$ along $\boldsymbol{d}_\mathrm{OS}$ in the time $t_\mathrm{u}$.} 
 \label{fig:time_delay}
\end{figure}

The second part of the right hand side of Equation~\eqref{eq:t}, $t_\mathrm{g}$, accounts for the additional path length to be travelled from the source position to the observer due to the deflection in the lens plane.
Figure~\ref{fig:time_delay} sketches the the situation. 
From it, we can derive $t_\mathrm{g}$ as 
\begin{equation}
t_\mathrm{g} = \dfrac{\left| \boldsymbol{d}_\mathrm{OL} \right| + \left| \boldsymbol{d}_\mathrm{LS} \right| - \left| \boldsymbol{d}_\mathrm{OS} \right|}{c} \;,
\label{eq:time}
\end{equation}
in which the $\boldsymbol{d}_{IJ} \in \mathbb{R}^3$ denote the three dimensional vectors of the proper distances connecting the observer $O$, the deflected image $L$ in the lens plane, and the source $S$.
Approximating
\begin{align}
\left| \boldsymbol{d}_\mathrm{LS} \right| &= \sqrt{\boldsymbol{d}_\mathrm{OL}^2 + \boldsymbol{d}_\mathrm{OS}^2 - 2 \boldsymbol{d}_\mathrm{OL} \cdot \boldsymbol{d}_\mathrm{OS}}\\
 &\approx  \left| \boldsymbol{d}_\mathrm{OS} \right| - \left| \boldsymbol{d}_\mathrm{OL} \right| + \dfrac{ \left| \boldsymbol{d}_\mathrm{OS} \right|  \left| \boldsymbol{d}_\mathrm{OL} \right|}{ \left| \boldsymbol{d}_\mathrm{OS} \right| -  \left| \boldsymbol{d}_\mathrm{OL} \right|} \dfrac{\boldsymbol{\alpha}(\boldsymbol{x})^2}{2}
\end{align}
and inserting it into Equation~\eqref{eq:time}, we arrive at 
\begin{equation}
t_g \approx \dfrac{ \left| \boldsymbol{d}_\mathrm{OS} \right|  \left| \boldsymbol{d}_\mathrm{OL} \right|}{ \left| \boldsymbol{d}_\mathrm{LS} \right|} \dfrac{\boldsymbol{\alpha}(\boldsymbol{x})^2}{2c}  = \dfrac{ \left| \boldsymbol{d}_\mathrm{OS} \right|  \left| \boldsymbol{d}_\mathrm{OL} \right|}{ \left| \boldsymbol{d}_\mathrm{LS} \right|} \dfrac{\left(\boldsymbol{x} - \boldsymbol{y} \right)^2}{2c} \;,
\label{eq:t_g}
\end{equation}
as derived in more detail in \cite{bib:Wagner6}\footnote{While \cite{bib:Wagner6} treated the specific case of a gravitational deflection potential, the derivation of the arrival time difference in a Friedmann cosmology also holds independent of the cause of the deviation from the unperturbed path.}.
Converting the proper distances to angular diameter distances by scaling with the redshifts, we obtain for the difference of arrival times between a perturbed and an unperturbed light ray
\begin{align}
t_\mathrm{g} &= \dfrac{(1+z_\mathrm{d})}{2c} \dfrac{D_\mathrm{d} D_\mathrm{s}}{D_\mathrm{ds}} \boldsymbol{\alpha}(\boldsymbol{x})^2 =  \dfrac{(1+z_\mathrm{d})}{2c} \dfrac{D_\mathrm{d} D_\mathrm{s}}{D_\mathrm{ds}} \left( \boldsymbol{x} - \boldsymbol{y} \right)^2 \;.  \label{eq:tg1}
\end{align}

Next, we relate $\tilde{\psi}(\boldsymbol{x})$ to $\psi(\boldsymbol{x})$ by Fermat's principle, i.e.~the extrema of the arrival time delay difference function should be the points at which the deflected images occur.
Setting $\nabla_{\boldsymbol{x}} t$ to zero, we thus obtain for the relation between $\tilde{\psi}(\boldsymbol{x})$ and $\psi(\boldsymbol{x})$
\begin{equation}
\psi(\boldsymbol{x}) = \dfrac{D_\mathrm{ds}}{D_\mathrm{d}D_\mathrm{s}} \tilde{\psi}(\boldsymbol{x}) \;.
\label{eq:psi}
\end{equation}

Adding Equations~\eqref{eq:t_psi} and \eqref{eq:tg1}, we finally obtain
\begin{align}
t &= \dfrac{(1+z_\mathrm{d})}{c} \dfrac{D_\mathrm{d} D_\mathrm{s}}{D_\mathrm{ds}} \left( \dfrac12 \boldsymbol{\alpha}(\boldsymbol{x})^2 -  \psi(\boldsymbol{x}) \right) \equiv \Gamma \left( \dfrac12 \boldsymbol{\alpha}(\boldsymbol{x})^2 -  \psi(\boldsymbol{x}) \right) \label{eq:time_delay0}\\
&= \Gamma \left( \dfrac12 \left( \boldsymbol{x} - \boldsymbol{y} \right)^2 - \psi(\boldsymbol{x}) \right). 
\label{eq:time_delay1}
\end{align}
The difference in arrival times can thus be separated into two parts.
The first one contains the geometric configuration of the observer, the lens, and the source, $\Gamma$, and is dependent on the underlying cosmology for the light propagation from the source to the observer.
The second part, the term in brackets, only contains dimensionless quantities and hence describes the actual effective light deflection.

As we do not know the emission time of a light signal at the source position, we cannot directly measure the difference in travel time between an unperturbed light ray and a perturbed one. 
Instead, we need to observe the time difference between two light pulses of a periodically time-varying source with and without a lens between the source and the observer.

Assuming that the source cannot be observed, we have to eliminate it from the equations. 
This is only possible, if observations from at least two images coming from the same source are available.
Eliminating the source out of Equation~\eqref{eq:lens_equation2} using the image positions from a pair of images at positions $\boldsymbol{x}_1$ and $\boldsymbol{x}_2$, sharing the same source at $\boldsymbol{y}$, we obtain 
\begin{equation}
\boldsymbol{x}_1 - \boldsymbol{x}_2 = \boldsymbol{\alpha}(\boldsymbol{x}_1) - \boldsymbol{\alpha}(\boldsymbol{x}_2) \;.
\label{eq:lens_equation3}
\end{equation}
This is one equation with two unknowns, so that further constraining equations are necessary to solve for the deflection angles.
Yet, adding further multiple images from the same source, for each additional equation, an additional unknown is added, such that the system of equations remains under-constrained without employing further constraints, e.g.~from an observed symmetry in the lens.

\begin{figure*}[t!]
\centering
\includegraphics[width=0.45\textwidth]{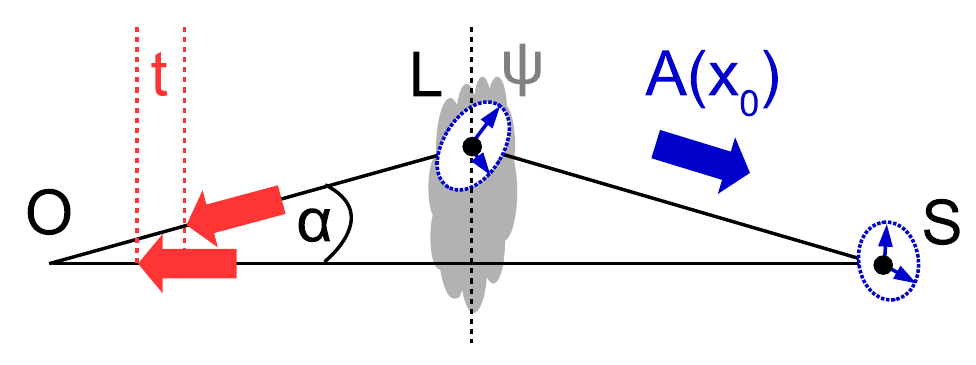} \hspace{6ex}
\includegraphics[width=0.45\textwidth]{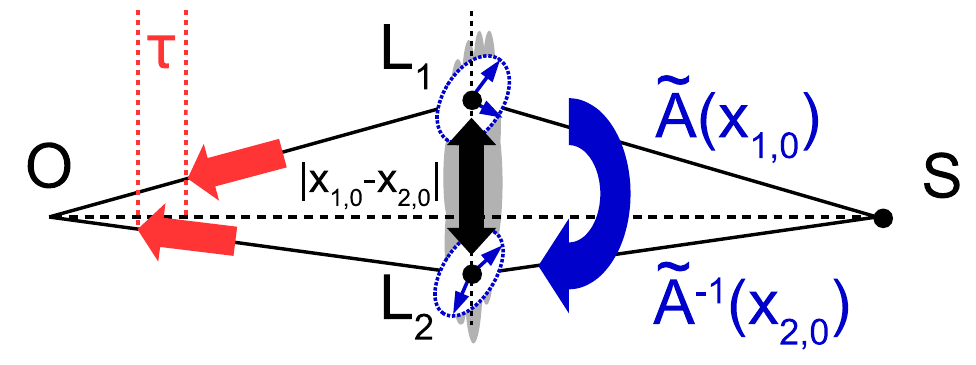} 
\caption{Summary of observables and inferred lens properties for an observable source (left) and an occluded source (right). 
Left: For an observable, periodically time-varying source, the deflection potential $\psi(\boldsymbol{x}_0)$ at the image position $\boldsymbol{x}_0$ can be determined from the arrival time difference $t$ between the perturbed and the unperturbed light ray, given $\boldsymbol{x}_0$ and the source position $\boldsymbol{y}_0$. The deflection angle $\boldsymbol{\alpha}(\boldsymbol{x}_0)$ can be determined by subtracting $\boldsymbol{y}_0$ from $\boldsymbol{x}_0$. From the shape of a potentially extended source and its image, the distortion matrix $A(\boldsymbol{x}_0)$ can be determined. 
Right: If the source is not observable, only the entries of $n$ distortion matrices up to an overall scaling factor, $\tilde{A}(\boldsymbol{x}_{i,0})$, $i=1,2,..., n$, can be determined for at least three extended multiple images that contain at least two non-parallel vectors in each image which can be mapped onto each other. The relative distance between a pair of multiple images only constrains the difference between their deflection angles, $\boldsymbol{\alpha}(\boldsymbol{x}_{1,0}) - \boldsymbol{\alpha}(\boldsymbol{x}_{2,0})$. The arrival time difference between a pair of images from the same source can be employed to infer the combination of deflection angles and potential differences according to Equations~\eqref{eq:time_delay2} and \eqref{eq:time_delay3}. If a common source position $\boldsymbol{y}_0$ can be found by the back-projection of the multiple images into the source plane using additional assumptions about the lens, the deflection angles at the image positions and the differences of the potential values can be self-consistently determined.} 
 \label{fig:observables}
\end{figure*}

Analogously, we can eliminate the vector in the source plane $\boldsymbol{y} - \boldsymbol{y}_0$ from Equation~\eqref{eq:A}.
Contrary to Equation~\eqref{eq:lens_equation3}, it becomes possible to determine the entries of the distortion matrix up to an overall scaling factor, as further detailed in \cite{bib:Wagner2}.
Employing at least two non-parallel vectors in at least three images from the same source, we can solve for $\tilde{A}(\boldsymbol{x}_{i,0})$, $i=1,2,...$, i.e.~the distortion matrices up to an overall scale factor at the positions of the multiple images $\boldsymbol{x}_{i,0}$, $i=1,2,...$, around which the distortion matrices are constant to a good approximation.

To eliminate the unknown, unperturbed path in Equation~\eqref{eq:time_delay1}, we determine the arrival time difference between two images coming from the same source as
\begin{align}
\tau &= \Gamma \left( \dfrac{(\boldsymbol{x}_1-\boldsymbol{x}_2)(\boldsymbol{\alpha}(\boldsymbol{x}_1)+\boldsymbol{\alpha}(\boldsymbol{x}_2))}{2} - (\psi(\boldsymbol{x}_1) - \psi(\boldsymbol{x}_2) ) \right) \label{eq:time_delay2} \\
&= \Gamma \left( \dfrac{(\boldsymbol{x}_1-\boldsymbol{y})^2 - (\boldsymbol{x}_2 - \boldsymbol{y})^2}{2} -(\psi(\boldsymbol{x}_1) - \psi(\boldsymbol{x}_2) )  \right) \;.
\label{eq:time_delay3}
\end{align}
Since the two images travel different paths through the deflecting structure, they appear at two different positions, $\boldsymbol{x}_1$ and $\boldsymbol{x}_2$ in the lens plane. 
Consequently, they are subject to the local values of the deflection potential at these positions, as well as the local values of the deflection angle at these points. 
Neither the deflection angles nor the source position are known, so that they have to be self-consistently inferred from overlapping back-projections of the multiple images.
The only model-independent statement the arrival time difference can make, is about the relationship between the absolute value of the average gradient 
\begin{equation}
\mathcal{G}(\boldsymbol{x}_1,\boldsymbol{x}_2,\psi) \equiv \left| \dfrac{1}{2} \Big( \boldsymbol{\alpha}(\boldsymbol{x}_1)+\boldsymbol{\alpha}(\boldsymbol{x}_2) \Big) \right| = \left| \dfrac{1}{2} \nabla \Big( \psi(\boldsymbol{x}_1)+\psi(\boldsymbol{x}_2) \Big) \right|
\end{equation}
and the difference quotient of the deflecting potential at the two image positions
\begin{equation}
\mathcal{D}(\boldsymbol{x}_1,\boldsymbol{x}_2,\psi) \equiv \dfrac{\left| \psi(\boldsymbol{x}_1) - \psi(\boldsymbol{x}_2) \right|}{\left| \boldsymbol{x}_1 - \boldsymbol{x}_2  \right|}\;.
\end{equation}
By the mean value theorem, 
\begin{equation}
\mathcal{D}(\boldsymbol{x}_1,\boldsymbol{x}_2,\psi) \le  \mathcal{G}(\boldsymbol{x}_1,\boldsymbol{x}_2,\psi) \;.
\label{eq:mvt} 
\end{equation}
the term in the large brackets in Equation~\eqref{eq:time_delay2} only vanishes for a linear deflection potential, for which equality holds in Equation~\eqref{eq:mvt}. For a fixed non-zero distance between the multiple image pair, the arrival time difference can thus be considered as a measure of local non-linearities for the deflection potential. 
Figure~\ref{fig:observables} summarises the observables and the lens properties that can be inferred for an observable source (left) and an occluded source (right).

\section{Gravitational and plasma lensing}
\label{sec:differences}

Given the derivations of Section~\ref{sec:formalism}, we apply these general formulae to the cases of gravitational and plasma lensing. 
Section~\ref{sec:gravitation} briefly treats the case of deflecting mass densities.
A more detailed derivation can be found in the books of \cite{bib:Petters} and \cite{bib:SEF} or in the paper \cite{bib:Wagner6}.
Section~\ref{sec:plasma} subsequently elaborates the case of plasma lensing. 
The derivations mainly rely on those of \cite{bib:Cordes}, \cite{bib:Grillo}, \cite{bib:Rogers}, and \cite{bib:Tuntsov}.
We unify the approaches and show that their definitions are consistent with each other.
In the end, Table~\ref{tab:comparison} in Section~\ref{sec:conclusion} summarises the results of this section.

\subsection{Gravitational lensing}
\label{sec:gravitation}

\subsubsection{Deflection potential}
\label{sec:potential_mass}

Based on the formalism derived in Section~\ref{sec:formalism}, we now have to specify the deflection potential for a gravitational lens, project it into the lens plane and scale it to obtain the dimensionless, projected deflection potential of Equation~\eqref{eq:psi}.
As derived in detail in \cite{bib:Wagner6}, we obtain for the projected dimensionless deflection potential
\begin{equation}
\psi_\mathrm{m}(\boldsymbol{x}) = \dfrac{D_\mathrm{ds}}{D_\mathrm{d}D_\mathrm{s}} \tilde{\psi}_\mathrm{m}(\boldsymbol{x}) = \dfrac{D_\mathrm{ds}}{D_\mathrm{d}D_\mathrm{s}} \dfrac{2}{c^2} \int \limits_{|\boldsymbol{d}_\mathrm{OS}|}^{0} \mathrm{d} l \, \phi_\mathrm{m}(\boldsymbol{r}) \;,
\label{eq:psi_m}
\end{equation}
in which the integral is the projection along the line of sight of the three-dimensional Newtonian gravitational potential $\phi_\mathrm{m}(\boldsymbol{r})$. 
The latter is sourced by a deflecting mass density $\rho(\boldsymbol{r})$, which is assumed to be of finite extent around the lens plane. 
The relationship between $\phi_\mathrm{m}(\boldsymbol{r})$ and $\rho(\boldsymbol{r})$ is given by the three-dimensional Poisson equation 
\begin{equation}
\Delta_{\boldsymbol{r}} \phi_\mathrm{m}(\boldsymbol{r}) =  4 \pi G \rho(\boldsymbol{r}) \;.
\label{eq:poisson_3d}
\end{equation}
The potential fulfils the conditions that it is negative for all $\boldsymbol{r}$ inside the deflecting mass and that $\phi_\mathrm{m}(\boldsymbol{r}) \rightarrow 0$ at its boundary\footnote{See \cite{bib:Wagner4} for further details about the boundary conditions that can be employed for deflecting masses of finite and infinite extent.}.
Being negative inside the deflecting mass implies that the refractive index is larger than 1, such that gravitation causes a retardation of light propagation with respect to the light propagation in vacuum.

Projecting the potential and the mass density into the lens plane, the Poisson equation becomes two-dimensional, relating $\psi_\mathrm{m}(\boldsymbol{x})$ to the two-dimensional mass density $\Sigma(\boldsymbol{x})$ by
\begin{equation}
\Delta_{\boldsymbol{x}} \psi_\mathrm{m}(\boldsymbol{x}) = 2 \dfrac{\Sigma(\boldsymbol{x})}{\Sigma_\mathrm{c}} \equiv 2 \kappa(\boldsymbol{x})\;,
\label{eq:poisson}
\end{equation}
in which $\kappa(\boldsymbol{x})$ is called the convergence.
It is physically reasonable, to assume that $\kappa(\boldsymbol{x})$ is a continuous, non-negative function with compact support as it represents the dimensionless projected deflecting mass density of finite extent.
Combining Equations~\eqref{eq:psi_m} and \eqref{eq:poisson} and assuming that the deflecting mass density extends over an area $\mathcal{A}$ in the lens plane yields
\begin{equation}
\psi_\mathrm{m}(\boldsymbol{x}) = \dfrac{1}{\pi} \int_{\mathcal{A}} \mathrm{d}^2 \tilde{\boldsymbol{x}} \, \kappa(\tilde{\boldsymbol{x}}) \ln \left| \boldsymbol{x} - \tilde{\boldsymbol{x}}\right| \;.
\label{eq:psi_kappa}
\end{equation}

\subsubsection{Point model of gravitational lensing}
\label{sec:point_mass}

The simplest surface mass density distribution in the lens plane at angular diameter distance $D_\mathrm{d}$ is a point mass $m$ at the origin such that its convergence is given by
\begin{equation}
\kappa(\boldsymbol{x}) =  \dfrac{\tfrac{m}{D_\mathrm{d}^2}\delta(\boldsymbol{x})}{\Sigma_\mathrm{c}} \;.
\label{eq:kappa_pm}
\end{equation}
Inserting Equation~\eqref{eq:kappa_pm} into Equation~\eqref{eq:psi_kappa}, we obtain
\begin{equation}
\psi_\mathrm{m}(\boldsymbol{x}) = \dfrac{4Gm}{c^2} \dfrac{D_\mathrm{ds}}{D_\mathrm{d} D_\mathrm{s}} \ln \left| \boldsymbol{x} \right| \;.
\label{eq:point_mass}
\end{equation}
Hence, the most fundamental element of a gravitationally deflecting structure is proportional to the Green's function of the two-dimensional Laplace operator.

\subsubsection{Gravitational lensing observables}
\label{sec:observables_mass}

In gravitational lensing configurations involving galaxy-scale and galaxy-cluster scale lenses, the sources are unobservable, so that the second set of formulae of Section~\ref{sec:formalism} has to be employed.
Known spectroscopic redshifts of the lens and the source can be inserted into the distance-redshift relation to determine the angular diameter distances in $\Gamma$.
This distance-redshift relation is usually based on a specific cosmological background model, yet, can also be set up based on observations, see \cite{bib:Wagner5} and references therein.

Multiple image configurations are usually corroborated by comparing the spectra of the images. 
If the images come from a common source with a time-varying intensity profile, we can measure the differences in arrival times between pairs of multiple images and insert them into the left-hand side of Equations~\eqref{eq:time_delay2} or \eqref{eq:time_delay3}.

The relative difference in the angular position of the multiple images can also be observed, either as the positions of point-like images or as the centres of light of extended images. 
They are required in the right-hand side of Equation~\eqref{eq:time_delay2} and in the approach detailed in \cite{bib:Wagner2}, \cite{bib:Wagner_cluster}.
If the multiple images have an extended intensity profile, the shape of the profile can be expressed either as a decomposition of the intensity profile into an orthonormal basis or as vectors between salient features within the intensity distribution (see \cite{bib:Wagner2}). 
The observed image morphologies can then be employed to characterise the local distorting properties of the lens and local approximations to the critical curves of the lens equation, as detailed in \cite{bib:Wagner_sum}.
The latter are independent of any specific mass density profile and are therefore robust lens-model-independent characterisations of the gravitational lens.

Gravitational lensing is achromatic, i.e.~independent of the wavelength at which observations are performed.
Consequently, any change in the multiple image configuration over a range of wavelengths is caused by different attenuation and absorption properties of the deflecting medium and by the changes in the emission profile of the source. 
Furthermore, the different wavelengths also allow to probe the deflecting structure on different scales, see e.g.~\cite{bib:Wagner_quasar} for an example.

Finally, as shown e.g.~in \cite{bib:Dyer}, the polarisation of multiple images is the same as the polarisation of their common source for static gravitational lenses to good approximation. Hence, observations of differences in the polarisation between multiple images can give insights into the effective magnetic field configuration of the lens.
In addition, as detailed in \cite{bib:Biggs} polarisation measurements can constrain time delay differences between pairs of multiple images.

\subsection{Plasma lensing}
\label{sec:plasma}

\subsubsection{Deflection potential}
\label{sec:potential_plasma}

Analogously to Section~\ref{sec:gravitation}, we derive the deflection potential for an inhomogeneous plasma lens from its effective refractive index, given the prerequisites of Section~\ref{sec:prerequisites}.
We further have to specify the plasma properties, because the refractive index not only depends on the frequency of the transient electro-magnetic signal, but also on the thermal velocity and the magnetisation of the ion gas. 
The refraction index of a three-dimensional, cold, unmagnetised plasma at position $\boldsymbol{r}$ for an electro-magnetic wave observed at an angular frequency $\omega$, $\omega=2\pi c/\lambda $ with the wave length $\lambda$, is given by
\be
n_{p}(\boldsymbol{r},\omega)= \sqrt{ 1- \dfrac{\omega_\mathrm{p}(\boldsymbol{r})^2}{\omega^2}} \approx 1- \dfrac{\omega_\mathrm{p}(\boldsymbol{r})^2}{2 \omega^2} \;,
\label{eq:n_plasma}
\ee
in which the plasma frequency is given by Equation~\eqref{eq:wp}, which depends on the position due to the changing electron density within the plasma.
$\omega \gg  \omega_\mathrm{p}$ is assumed for the approximation.
From Equation~\eqref{eq:n_plasma}, we read off that the phase space velocity of the wave traversing the plasma is larger than $c$ (the group velocity remains smaller than $c$), such that the arrival time difference in Equation~\eqref{eq:t_psi} is negative.
Rewriting $\omega_\mathrm{p}$ as given by the right-hand side of Equation~\eqref{eq:wp} and writing $\omega$ in terms of $\lambda$ yields
\begin{equation}
n_\mathrm{p}(\boldsymbol{r},\lambda) = 1 -  \dfrac{r_\mathrm{e} \lambda^2}{2\pi} n_\mathrm{e}(\boldsymbol{r}) \;,
\end{equation}
such that Equation~\eqref{eq:psi_phi} for plasma lensing reads
\begin{equation}
\tilde{\psi}_\mathrm{p}(\boldsymbol{x},\lambda) \equiv -  \dfrac{r_\mathrm{e} \lambda^2}{2\pi} \int \limits_{|\boldsymbol{d}_\mathrm{OS}|}^{0}  \mathrm{d} l \, n_\mathrm{e}(\boldsymbol{r}) 
\equiv  - \dfrac{r_\mathrm{e} \lambda^2}{2\pi} N_\mathrm{e}(\boldsymbol{x}) \;.
\end{equation}
The electron density $N_\mathrm{e}(\boldsymbol{x})$ which is obtained by integrating the three-dimensional electron density along the line of sight is often abbreviated as $\mathrm{DM}(\boldsymbol{x})$ and called the dispersion measure.

Contrary to the gravitational potential, the source of the deflection potential of the plasma, the electron density of the plasma, is directly proportional to the potential.
The physical reason for this difference is that the plasma potential is not the standard electric potential, which is the analogue of the gravitational potential in Equation~\eqref{eq:poisson_3d}.
Instead, the plasma potential causing the deflection is the scaled phase difference between the light travelling through the plasma and travelling through vacuum, as is shown, for instance, in \cite{bib:Cordes} and \cite{bib:Grillo}.

Solving the Poisson equation for the gravitational deflection potential, we arrive at Equation~\eqref{eq:psi_kappa} which shows that $\psi_\mathrm{m}(\boldsymbol{x})$ does not only contain information about the deflecting mass density at position $\boldsymbol{x}$, but also about all the non-local mass density contributions in the entire lensing region by means of the integration.
Contrary to that, $\psi_\mathrm{p}(\boldsymbol{x})$ only contains the local information about the electron density at $\boldsymbol{x}$, such that the lensing formalism for plasma lenses is a completely local approach, missing any mathematical regularity conditions that the Poisson equation implies (see \cite{bib:Wagner4} for details).

The deflection potential scaled according to Equation~\eqref{eq:psi} reads
\be
\psi_\mathrm{p}(\boldsymbol{x}) = - \dfrac{D_{ds}}{D_d D_s} {r_\mathrm{e} \lambda^2 \over 2\pi} N_e(\boldsymbol{x}) \equiv - \Gamma_\mathrm{p} N_e(\boldsymbol{x}) \;.
\label{eq:psi_plasma}
\ee
This implies that the total arrival time difference of Equation~\eqref{eq:time_delay1} can be written as
\begin{align}
t &= \Gamma \left( \dfrac{(\boldsymbol{x}-\boldsymbol{y})^2}{2} +  \Gamma_\mathrm{p} N_e(\boldsymbol{x}) \right) = \Gamma \, \Gamma_\mathrm{p} \left(\dfrac{\Gamma_\mathrm{p}}{2} \left( \nabla N_e(\boldsymbol{x})\right)^2 + N_e(\boldsymbol{x}) \right) \label{eq:t_plasma}
\\
&=  \dfrac{(1+z_\mathrm{d})}{c} \dfrac{r_\mathrm{e}\lambda^2}{2\pi}  \left(\dfrac{\Gamma_\mathrm{p}}{2} \left( \nabla N_e(\boldsymbol{x})\right)^2 + N_e(\boldsymbol{x}) \right) \;.
\end{align}
The lens equation, Equation~\eqref{eq:lens_equation1} reads
\begin{equation}
\boldsymbol{y} - \boldsymbol{x} = \Gamma_\mathrm{p} \nabla N_e(\boldsymbol{x}) \;.
\label{eq:alpha_plasma}
\end{equation}
Comparing Equations~\eqref{eq:psi_m} and \eqref{eq:psi_plasma}, we note that the deflection potentials have different signs which causes the light ray traversing the medium being retarded and advanced compared to the unperturbed light ray, respectively.
Consequently, the signs in front of the potential terms in the arrival time difference and in the lens equation are also opposite to each other.

Eliminating a potentially unobservable source, Equation~\eqref{eq:time_delay2} reads
\begin{equation}
\tau = \Gamma_\tau \left(-\dfrac{(\boldsymbol{x}_1-\boldsymbol{x}_2)}{2} \nabla \Big( N_\mathrm{e}(\boldsymbol{x}_1) + N_\mathrm{e}(\boldsymbol{x}_2)\Big) + N_\mathrm{e}(\boldsymbol{x}_1) - N_\mathrm{e}(\boldsymbol{x}_2) \right)
\label{eq:tau_plasma}
\end{equation} 
with the geometry-independent prefactor
\begin{equation}
\Gamma_\tau \equiv \Gamma \, \Gamma_\mathrm{p} = \dfrac{(1+z_\mathrm{d})}{c}\dfrac{r_\mathrm{e} \lambda^2}{2\pi} \;.
\end{equation}
Equation~\eqref{eq:lens_equation1} for an unobservable source yields
\begin{equation}
\boldsymbol{x}_1 - \boldsymbol{x}_2 = - \Gamma_\mathrm{p} \nabla \Big( N_\mathrm{e}(\boldsymbol{x}_1) - N_\mathrm{e}(\boldsymbol{x}_2) \Big) \;.
\label{eq:alpha_plasma2}
\end{equation}

\subsubsection{Point model of plasma lensing}
\label{sec:point_plasma}

Unlike gravitational lensing, we do not solve Poisson equation to obtain the lensing potential from a given electron density distribution. 
Therefore, the point model of plasma lensing is different from Equation~\eqref{eq:point_mass}. 
It is a negative point charge at the origin, such that the projected electron density of a point plasma lens is given by
\be
N_e(\boldsymbol{x}) = -\, \delta(\boldsymbol{x}) \;.
\ee
Its deflection potential is positive and reads
\be
\psi_\mathrm{p}(\boldsymbol{x})  = - \dfrac{D_{ds}}{D_d D_s} {r_\mathrm{e} \lambda^2 \over 2\pi} (-\delta(\boldsymbol{x})) = \Gamma_\mathrm{p} \delta(\boldsymbol{x}) \;.
\label{eq:point_plasma}
\ee
Since $t_\psi = t_\mathrm{p} - t_\mathrm{u} \propto -\psi_\mathrm{p}$ according to Equation~\eqref{eq:t_psi}, this potential causes an advanced phase in the perturbed light with respect to the unperturbed, which we expect because the refraction index of plasma lensing with respect to vacuum should always be smaller than one.
Defining the two-dimensional $\delta$-distribution as the limit of a Gaussian distribution with vanishing width $\sigma$
\begin{equation}
\delta(\boldsymbol{x}) = \lim \limits_{\sigma \rightarrow 0} \, \dfrac{1}{2\pi \sigma^2} \exp \left(- \dfrac{\boldsymbol{x}^2}{2\sigma^2} \right)
\label{eq:Gauss_lens}
\end{equation}
we can motivate the first plasma lens models, e.g.~the one-dimensional version of \cite{bib:Clegg}, as being the most simple, finite plasma lens models.
They can serve as the building blocks for more complex plasma lens configurations, similar to assembling a gravitational lens by point masses.
Such plasma Gaussian mixture models have already been successfully matched to describe observational plasma lensing configurations, see e.g.~\cite{bib:Cognard}.
The deflection angle derived from Equation~\eqref{eq:point_plasma} using Equation~\eqref{eq:Gauss_lens} reads
\be
\boldsymbol{\alpha}_\mathrm{p}(\boldsymbol{x})=\nabla_{\boldsymbol{x}} \psi_\mathrm{p} (\boldsymbol{x}) = \lim\limits_{\sigma\to 0} \, \dfrac{\Gamma_\mathrm{p}}{2\pi \sigma^2}  \left( -  \dfrac{\boldsymbol{x}}{\sigma^2}  \right) \exp\rund{-\dfrac{\boldsymbol{x}^2}{2\sigma^2}} \;.
\ee
The deflection angle approaches (0,0) for $\boldsymbol{x} \ne (0,0)$, but diverges when $\boldsymbol{x} \to (0,0)$.
Consequently, Equation~\eqref{eq:time_delay0} is
\be
t=\lim\limits_{\sigma\to 0} \, \Gamma \dfrac{\Gamma_\mathrm{p}}{2\pi \sigma^2} \exp \left( -\dfrac{\boldsymbol{x}^2}{2\sigma^2} \right) \left( \dfrac{\boldsymbol{x}^ 2}{2 \sigma^4} \dfrac{\Gamma_\mathrm{p}}{2\pi \sigma^2} \exp \left( -\dfrac{\boldsymbol{x}^2}{2\sigma^2} \right) -1 \right) \;.
\ee
The time delay is also 0 in the limit of $\sigma\to 0$ for $\boldsymbol{x} \ne (0,0)$, but diverges for $\boldsymbol{x}\to (0,0)$.
It is important to stress that this equation is obtained by inserting a positive potential into Equation~\eqref{eq:time_delay0}, such that $t = t_\mathrm{p} - t_\mathrm{u}$ can consistently cause an advanced light propagation for the case that $t_\mathrm{g} < t_\psi$.

Hence, the behaviour of this Gaussian plasma lens in the limit of a point plasma lens is in agreement with physical expectations:
As for a point charge with low temperature, we neglect the energy exchange between the charged particle and the light. 
The light will not interact with the charged particle unless it collides with the point charge at $\boldsymbol{x}=(0,0)$, where both the absolute value of the deflection angle and the arrival time difference between the perturbed and unperturbed light rays are infinite.

\subsubsection{Plasma lensing observables}
\label{sec:observables_plasma}

Since plasma lensing sources are often unresolved and time-varying, e.g. pulsars or fast radio bursts, the main observables are arrival time differences and the flux densities at point-like or scatter-broadened image positions in the lens plane. 
As a great benefit, plasma lensing is wavelength-dependent, so that these observables are usually available for an entire range of wavelengths.
The latter is typically between a few hundred MHz ($\approx 300$~MHz) and a few GHz ($\approx 10$~GHz) binned with a resolution of a few MHz ($\approx 4$~MHz), see e.g.~\cite{bib:Bannister} or \cite{bib:Tuntsov}.

Observations of flux density measurements over a period of time for a range of wavelengths are also available, see e.g.~\cite{bib:Walker}, and may yield valuable information about symmetries in the lens.
We focus on static lensing configurations at one instant in observing time and therefore defer the usage of changes in observed flux density over time for future work. 

Equations~\eqref{eq:t_plasma} and \eqref{eq:alpha_plasma} contain distances to the lens and the source and relative to each other. 
As we may not always have this information, we need to analyse the knowledge gain from these equations for known and unknown geometry, as detailed in Sections~\ref{sec:degeneracies} and \ref{sec:multi_wavelengths}.

Depending on the location of the plasma lens and the source along the line of sight, different distance measures may be available to determine $D_\mathrm{d}$, $D_\mathrm{s}$, and $D_\mathrm{ds}$. 
For configurations at extra-galactic distances, like those of fast radio bursts with measured redshift, a distance-redshift relation can employed like the standard one based on a cosmological model, like, for instance given by Equation~\eqref{eq:D_A}.
For configurations in our cosmic neighbourhood, observation-based distance measures may be suitable, as detailed in \cite{bib:Cordes2}.
If the plasma lens is in our Galaxy, we can also set $z_\mathrm{d} \approx 0$ to good approximation. 
The method to obtain the distances is important as shown in Section~\ref{sec:multi_wavelengths} because, in order to break the formalism-intrinsic degeneracies, the distance measure should be independent of $\boldsymbol{x}$, and independent of the observation wavelength.

The deflection angles of plasma lenses were derived in \cite{bib:Clegg}, \cite{bib:Er1}, \cite{bib:Er2}, and \cite{bib:Er4}.
These works found that the deflection angles are of the order of milli-arcseconds (mas) or even smaller. 
They thus require a high astrometric precision to be observed, as we will also show in our simulation in Section~\ref{sec:example}.
For fast radio bursts, this localisation precision is still challenging, but recent advancements in VLBI observations by \cite{bib:Marcote1}, \cite{bib:Marcote2} achieved a few mas localisation precision. 

In order to employ the local lens characterisation approach of \cite{bib:Wagner2} and \cite{bib:Wagner_cluster}, extended multiple images must be observed. Since, up to our knowledge so far, no cases suitable to apply this approach have been found, we defer further investigations in this direction to future work, as well. 

\subsection{Limits of the approach}
\label{sec:limits}

Assuming a quasi-stationary lens during the observations may not be fulfilled for a turbulent plasma. Furthermore, just like in microlensing, the velocity of a transversally moving cloud also yields information about the lens, which we will consider in future work.

\section{Degeneracies}
\label{sec:degeneracies}

This section contains an encompassing analysis of all potential degeneracies of gravitational and plasma lensing in the formalism as outlined in Section~\ref{sec:formalism}.
We first fix the degrees of freedom that belong to physically irrelevant invariance transforms like setting the coordinate system (Section~\ref{sec:optical_axis}) and the reference value for the refractive index (Section~\ref{sec:overdensities}).
Subsequently, we treat the degeneracies that imply physically different deflecting structures (Section~\ref{sec:plane_degeneracies}) or additionally altered geometries for the light propagation (Section~\ref{sec:los_degeneracies}).

\subsection{Invariance under a change of the optical axis}
\label{sec:optical_axis}

As detailed in \cite{bib:SEF}, the lens equation and the arrival time differences are invariant under a change of the optical axis.
In the formalism as outlined in Section~\ref{sec:formalism}, the optical axis connects the observer with the centre of the lens, such that the position of the source gets shifted from its true position. 
\cite{bib:SEF} show that the same equations as shown in Section~\ref{sec:formalism} arise based on the prerequisites of Section~\ref{sec:prerequisites} for any kind of deflecting structure, if the optical axis connects the lens and the source at the cost of shifting the observer's position from the position it had in the former case. 
The same result is also shown in \cite{bib:Grillo} for the case of a plasma lens as deflecting structure.

Since this degeneracy is merely a choice of a reference point, it is easily broken without any loss of generality for the physical phenomena to be described.
We consider it more natural to define the optical axis by connecting our position as the observer to the centre of the lens than defining the optical axis by connecting a possibly unobservable source position to the centre of the lens. 
Therefore, we keep breaking this degeneracy by the definition of the optical axis as done in Section~\ref{sec:formalism}.

\subsection{Over- and under-densities in the deflecting structure}
\label{sec:overdensities}

In a similar way as a reference point for the coordinate system has to be defined in Section~\ref{sec:optical_axis}, the deflecting potential also requires a reference value.
As defined by Equation~\eqref{eq:n}, this reference is the vacuum value, such that $\Sigma(\boldsymbol{x})$ of the gravitational lens and $N_\mathrm{e}(\boldsymbol{x})$ of the cold plasma lens are positive everywhere.
This is the physically most reasonable reference value because the refraction index for gravitational lensing of weak fields is derived from a perturbation of a Minkowski metric and the refraction index for plasma lensing is based on the definition of the plasma frequency.  
The latter sets a time scale on which plasma electrons moving at thermal velocities disturb the equilibrium electron density distribution given by $n_\mathrm{e}(\boldsymbol{r})$.

Yet, in principle, these definitions can be changed and the refractive index can be defined with respect to any reference value, potentially generating over- and under-densities in the deflecting structure with respect to this reference value.
Appendix~\ref{app:shifting_n} shows that choosing another reference value adds a constant offset to the arrival time difference in Equation~\eqref{eq:time_delay0} and leaves Equations~\eqref{eq:lens_equation2}, \eqref{eq:time_delay2}, and \eqref{eq:time_delay3} invariant. 
We can thus conclude that the equations relevant for gravitational lensing work with any reference value for the deflection potential. 
As also noted in \cite{bib:Liesenborgs2}, which is also valid in the general geometric optics approach of Section~\ref{sec:formalism}, we could even choose different reference values for different sets of multiple images, as long as all observables of multiple images coming from the same source have the same reference. 

Measuring the arrival time difference in Equation~\eqref{eq:time_delay0}, the reference light ray is the one as observed without the intervening lens. 
Usually, we assume or observe that this light travels through vacuum.
But, as shown in Appendix~\ref{app:shifting_n}, Equation~\eqref{eq:time_delay0} can easily account for a background medium.
A detailed derivation of light propagation through plasma with an arbitrary reference value for the refractive index can be found in \cite{bib:Bisnovatyi1}.
For the sake of clarity and simplicity, we keep fixing the freedom to choose the reference point of the refractive medium by the vacuum refractive index. 

The freedom to choose a global reference value for the potential variable does not change the topology of the lens being under- or overdense with respect to their environment because the reference value cancels out in the difference between the lens potential and the potential of its environment.
Over- and underdense lenses with respect to their environment can occur for plasma and for gravitational lenses. 
The latter is usually denoted as gravitational lensing by voids, see \cite{bib:Chen} and references therein for a recent overview.

\subsection{Invariance under a change of the deflecting structure within the lens plane}
\label{sec:plane_degeneracies}

Investigating the physically relevant degeneracies, starting with the degeneracies within the lens plane, we assume that $\Gamma$, i.e.~the geometry of the background universe in which the light rays propagate, is known and fixed.
To derive the invariance transformations within the lens plane, we treat the cases of gravitational and plasma lensing separately because, for gravitational lensing, we have to take into account the Poisson equation relating the deflection potential and the mass density of the lens  (Equation~\eqref{eq:poisson}), while for plasma lensing, the deflection potential is directly proportional to the deflecting electron density (Equation~\eqref{eq:psi_plasma}).
Furthermore, there are plasma lensing cases in which the source position is observable, which is not possible in the gravitational lensing we consider.
An encompassing overview of all degeneracies and the respective invariance transformations of the equations describing gravitational lensing can be found in \cite{bib:Wagner4}.
We summarise the results as follows to compare them to the case of plasma lensing.

\subsubsection{Gravitational lensing}
\label{sec:degeneracies_gl}

Having measured time delay differences between pairs of multiple images, we can employ Equations~\eqref{eq:time_delay2} and \eqref{eq:time_delay3} to constrain differences in the Fermat potential of the gravitational lens between the image positions.
The Fermat potential is the term in brackets containing the deflection angles or source positions and the difference in the deflection potential.

As discovered in \cite{bib:Saha} and \cite{bib:Liesenborgs1} a degeneracy in the global reconstruction of the convergence map in the lensing region remains which allows to add or subtract mass densities of finite extent which do not overlap with the multiple images.
In \cite{bib:Wagner4}, this degeneracy is shown to occur because the arrival time difference between multiple images only fixes the convergence maps in the region between these images \emph{almost} everywhere. 
This mathematical result physically allows for additional point masses or even finitely extended masses to be added or subtracted, such that the degeneracies found in \cite{bib:Saha} and \cite{bib:Liesenborgs1} can be directly derived from the mathematical construction of the lensing formalism.
As further detailed in theory in \cite{bib:Wagner6} and shown in practice in \cite{bib:Meneghetti2}, knowing the cosmological background of the gravitational lens, the enclosed mass within the critical curve can be determined for a large number of multiple images spread around the critical curves because they tightly constrain the shape of the critical curves. 
A precise estimate of the total enclosed mass then limits the possibilities to add or subtract masses which are drastically further decreased by imposing certain smoothness or shape constraints.  
In summary, measuring time delay differences is of limited usage to globally constrain the convergence in the entire lensing region. 
The gain in reconstruction accuracy is detailed in \cite{bib:Wagner_frb} and \cite{bib:Liesenborgs3}.

At least, time delay difference measurements fix the local deflection potential at the image positions up to a linear function, so that the convergence at these image positions, being a second derivative of $\psi_\mathrm{m}(\boldsymbol{x})$, is fixed. 
However, the convergence values at these positions are not accessible by measuring the time delay difference. 
As Equation~\eqref{eq:mvt} implies, the amount of the time delay difference for given distance between the images only measures by how much their local deflection potential difference deviates from linearity.

In addition, at the positions of the multiple images, there is a degeneracy between intrinsic source properties and local lens properties: 
To infer local lens properties, i.e. ratios of the distortion matrix entries, we rely on the transformation of multiple images onto each other, as detailed in \cite{bib:Wagner2} and \cite{bib:Wagner_cluster}. 
While this method becomes accurate in the limit of point images and source, we can only determine differential local lens properties between pairs of multiple images, so that a minimum distortion effect that all multiple images have in common is degenerate with the intrinsic morphology of their common background source.

Consequently, it is not possible to determine absolute distortion characteristics of a gravitational lens unless we know the source.
At best, we can arrive at a self-consistent local lens reconstruction and source morphology.
Since, the freedom remains to alter the mass density distribution in the entire lens plane, even as a non-constant function of the position, we call this degeneracy the ``mass-sheet degeneracy''.

\subsubsection{Plasma lensing}
\label{sec:degeneracies_pl}

For plasma lensing, we now pursue an analogous analysis as carried out in \cite{bib:Wagner4} to obtain the invariance transformations of all relevant equations for one fixed wavelength and give them a physical interpretation.
A lens reconstruction employing the equations in Section~\ref{sec:formulae} determines $N_\mathrm{e}(\boldsymbol{x})$ or its gradient only locally at the multiple image positions.
Due to this locality, we have the freedom to alter the electron density distribution in the entire lens plane except at the multiple image positions in a completely random way, as long as $\psi_\mathrm{p}$ remains differentiable at the multiple image positions.

To determine the invariance transformations at the positions of the multiple images, we first investigate plasma lensing configurations with known source position, i.e. Equations~\eqref{eq:t_plasma} and \eqref{eq:alpha_plasma}.
For a fixed geometrical configuration $\Gamma$, there is no degeneracy arising. 
If we can measure the light travel time through the background by observing two subsequent signals of a periodically radiating source and if we can measure the light travel time of two signals through the plasma lens, the arrival time difference immediately gives us $N_\mathrm{e}(\boldsymbol{x})$ by Equation~\eqref{eq:t_plasma}.
Analogously, we obtain $\nabla N_\mathrm{e}(\boldsymbol{x})$ from Equation~\eqref{eq:alpha_plasma} using observations of the source and the image position.

Next, we investigate plasma lensing configurations with unobservable source position, Equations~\eqref{eq:tau_plasma} and \eqref{eq:alpha_plasma2}.
Being derived from the general Equations~\eqref{eq:time_delay2} and \eqref{eq:lens_equation3} like the gravitational lensing configurations, we encounter the same degeneracies for plasma lenses.

To find all transformations of plasma lensing variables that leave the observables invariant, we locally perturb the electron density by
\begin{equation}
\delta N_\mathrm{e}(\boldsymbol{x}) \equiv \tilde{N}_\mathrm{e}(\boldsymbol{x})  - N_\mathrm{e}(\boldsymbol{x})  \;,
\end{equation}
then, leaving $\tau$ of Equation~\eqref{eq:tau_plasma} invariant by this change in the electron density, $\delta N_\mathrm{e}(\boldsymbol{x})$ must obey
\begin{equation}
\delta N_\mathrm{e}(\boldsymbol{x}_1) - \delta N_\mathrm{e}(\boldsymbol{x}_2) = \dfrac{(\boldsymbol{x}_1 -  \boldsymbol{x}_2)}{2 } \nabla \Big( \delta N_\mathrm{e}(\boldsymbol{x}_1) + \delta N_\mathrm{e}(\boldsymbol{x}_2) \Big) \;.
\label{eq:constr_tau}
\end{equation}
Hence, $\delta N_\mathrm{e}(\boldsymbol{x})$ has to be a linear function. 
Analogously, we require that the change in the electron density does not alter the observable distance between two multiple images in Equation~\eqref{eq:alpha_plasma2}, but, that we can change the unobservable source position by
\begin{equation}
\delta \boldsymbol{y} \equiv \tilde{\boldsymbol{y}} - \boldsymbol{y} \;,
\end{equation}
to find the constraint which is valid for all observed multiple images $i$
\begin{equation}
\delta \boldsymbol{y} = \Gamma_\mathrm{p} \nabla \delta N_\mathrm{e}(\boldsymbol{x}_i) \;.
\label{eq:constr_alpha}
\end{equation}
This is the same degeneracy also found for gravitational lenses, as detailed in \cite{bib:Wagner4}.
The constraints of Equations~\eqref{eq:constr_tau} and \eqref{eq:constr_alpha} together yield that $\delta N_\mathrm{e}(\boldsymbol{x})$ is a linear function and its gradient is a constant. 
Consequently, even if all multiple images can be back-projected to the source plane to have a perfect overlap with each other, the freedom remains to translate this reconstructed source in the source plane without changing the measured arrival time difference or the positions of the multiple images.
Assuming we can observe extended multiple images to infer local lens properties according to \cite{bib:Wagner2} and \cite{bib:Wagner_cluster}, we additionally encounter the degeneracy between the local lens and intrinsic source properties mentioned in Section~\ref{sec:degeneracies_gl}.

Compared to gravitational lensing, we thus arrive at the same physically relevant degeneracies for the case of unobservable sources, which we analogously dub the ``gas-sheet degeneracy''. 

\subsection{Invariance under a change of the deflecting structure along the line of sight}
\label{sec:los_degeneracies}

Having characterised the degeneracies in the deflecting structures within the lens plane, we now investigate the degeneracy between the geometric distance factor $\Gamma$ and the deflecting structure in the lens plane. 
For gravitational lenses, as detailed in \cite{bib:Wagner6}, this degeneracy arises because we can only measure the redshift of the lens and the source and the corresponding distances required in $\Gamma$ are calculated by means of a cosmological background model. 
The distance-redshift relation of this background model usually requires a cosmological background mass density to be fixed, on top of which the deflecting mass density is set up. 
Depending on the definition of the distance-redshift relation, we thus face the degeneracy between masses assigned to the cosmological background density and masses assigned to the projected deflecting structure. 
Employing the distance-redshift relation of the current cosmological concordance model to determine the angular diameter distance between two redshifts $z_1$ and $z_2$ with $z_1 < z_2$
\begin{equation}
D_\mathrm{A}(z_1,z_2) = \dfrac{c}{H_0} \dfrac{z_1+1}{z_2+1} \int \limits_{z_1}^{z_2} \dfrac{\mathrm{d}z}{E(z)}
\label{eq:D_A}
\end{equation}
we find that the degeneracy between the distance-redshift relation and the projected deflecting mass density is mainly sensitive to a rescaling of the Hubble constant $H_0$ and a corresponding rescaling of the deflection potential in Equation~\eqref{eq:psi_m}.
The terms representing the background mass density and potential forms of dark energy in the cosmic expansion function $E(z)$ only have a smaller contribution to this degeneracy, see e.g. \cite{bib:Wong}.
 
For plasma lenses, an analogous degeneracy only arises for the case of an observable source because, for an unobservable source, Equation~\eqref{eq:tau_plasma} is obviously independent of the light path geometry.
Transforming $\Gamma$ as $\delta \Gamma = \tilde{\Gamma} - \Gamma$ and transforming $N_\mathrm{e}(\boldsymbol{x})$ as $\delta N_\mathrm{e}(\boldsymbol{x}) = \tilde{N}_\mathrm{e}(\boldsymbol{x}) - N_\mathrm{e}(\boldsymbol{x})$, we can insert the transformed quantities into Equation~\eqref{eq:t_plasma}. 
Requiring that the transformation leaves $\tau$ invariant, we obtain the relation
\begin{equation}
\delta N_\mathrm{e}(\boldsymbol{x}) = - \dfrac{\left( \boldsymbol{x} - \boldsymbol{y} \right)^2}{2} \Gamma_\tau^{-1} \, \delta \Gamma \;.
\end{equation}
Deriving $\delta N_\mathrm{e}(\boldsymbol{x})$ yields
\begin{equation}
\nabla \delta N_\mathrm{e}(\boldsymbol{x}) = - \left( \boldsymbol{x} - \boldsymbol{y} \right) \Gamma_\tau^{-1} \, \delta \Gamma - \dfrac{\left( \boldsymbol{x} - \boldsymbol{y} \right)^2}{2} \Gamma_\tau^{-1}  \nabla\delta \Gamma \;.
\label{eq:derivative_deltaNe}
\end{equation}
Comparing Equation~\eqref{eq:derivative_deltaNe} with the transformation of Equation~\eqref{eq:alpha_plasma} that is supposed to leave $\boldsymbol{y} - \boldsymbol{x}$ as observable invariant, 
\begin{equation}
\dfrac{\nabla \tilde{N}_\mathrm{e}(\boldsymbol{x})}{\tilde{\Gamma}} = \dfrac{\nabla N_\mathrm{e}(\boldsymbol{x})}{\Gamma} \quad \Leftrightarrow \quad \nabla \delta N_\mathrm{e}(\boldsymbol{x}) = - \left( \boldsymbol{x} - \boldsymbol{y} \right)\Gamma_\tau^{-1} \, \delta \Gamma \;, 
\label{eq:los_degeneracy}
\end{equation}
we can conclude that $\delta \Gamma$ is a constant, as expected by construction, assuming that the distances do not depend on any lens plane position in the electron cloud.

Physically, the degeneracy between the distance ratio in $\Gamma$ and the electron density can be understood when we measure the distances to the luminous source and lens objects in terms of their distance modulus $\mu$.
While we usually employ angular diameter distances in the gravitational lensing formalism, any distance measure can be used. For instance, the distance to point-like sources like supernovae is usually determined by means of the distance modulus.
Then, the distance is given in terms of the apparent magnitude $m_B$ of the observed object and its absolute magnitude $M_B$ in some band $B$
\begin{equation}
D_\mathrm{L}(\mu) = 10^{\mu/5+1} = 10^{1-M_B/5} \cdot 10^{m_B/5} \;.
\end{equation}
By this formula, $D_\mathrm{L}$ is measured in parsecs. 
Not knowing the absolute magnitudes of the plasma lens or the source thus causes an analogous degeneracy as for the gravitational lenses.
While, for the gravitational lenses, the degeneracy refers to the distribution of the mass density between background and deflector, for plasma lenses the degeneracy means that we have the freedom to partition the observed luminosity between lens, source, and may even have to account for potential extinction along the path. 

\section{Breaking the plasma lensing degeneracies}
\label{sec:multi_wavelengths}

Since ways to break the degeneracies occurring for gravitational lensing configurations are detailed in \cite{bib:Wagner4} and \cite{bib:Wagner6}, we focus on breaking the degeneracies for plasma lensing configurations here. 
Table~\ref{tab:degeneracies} summarises the degeneracies found in Section~\ref{sec:degeneracies}.

\begin{table*}[t]
 \caption{Synopsis of all degeneracies determined in Section~\ref{sec:degeneracies}.}
\label{tab:degeneracies}
\begin{center}
\begin{tabular}{cccc}
\hline
\noalign{\smallskip}
 & \textbf{Geometry known} & \textbf{Geometry unknown} & \textbf{Degeneracy breaking} \\
 & (Section~\ref{sec:plane_degeneracies}) & (Section~\ref{sec:los_degeneracies}) &  (Section~\ref{sec:multi_wavelengths})\\ 
\noalign{\smallskip}
\hline
\noalign{\smallskip}
\textbf{Source known}  &  no & $\delta N_\mathrm{e}(\boldsymbol{x}) = - \tfrac{\left(\boldsymbol{x} -\boldsymbol{y}\right)^2}{2} \Gamma_\tau^{-1} \delta \Gamma$  & $\Gamma = \dfrac{\left(\delta t_+ + \delta t_- \right)}{\left(\delta \boldsymbol{x}_+ + \delta \boldsymbol{x}_- \right)\left(\boldsymbol{x}-\boldsymbol{y}\right)}$\\
\noalign{\smallskip}
& degeneracies & $\delta \Gamma = \text{const.}$ & (observations at $\lambda \pm \delta \lambda$, $\delta \lambda / \lambda \ll 1$) \\ 
\noalign{\smallskip}
\hline
\noalign{\smallskip}
 & $\delta N_\mathrm{e}(\boldsymbol{x}) = \boldsymbol{\xi_1} \boldsymbol{x} + \xi_2$ & $\delta N_\mathrm{e}(\boldsymbol{x}) = \boldsymbol{\xi_1} \boldsymbol{x} + \xi_2$  & \\
\textbf{Source unknown}   & $\boldsymbol{\xi}_1 \in \mathbb{R}^2, \xi_2 \in \mathbb{R}$ & $\boldsymbol{\xi}_1 \in \mathbb{R}^2, \xi_2 \in \mathbb{R}$ & $\tfrac{1}{2} \nabla (N_\mathrm{e}(\boldsymbol{x}_1)+N_\mathrm{e}(\boldsymbol{x}_2)) = - \tfrac{2\pi}{r_\mathrm{e} \lambda^2} \tfrac{c d^{-1}}{(1+z_\mathrm{d})}  \left( \begin{matrix} \delta \boldsymbol{x}_{-,2} & -\delta \boldsymbol{x}_{+,2} \\ -\delta \boldsymbol{x}_{-,1} & \delta \boldsymbol{x}_{+,1}\end{matrix} \right) \left( \begin{matrix} \delta \tau_+ \\ \delta \tau_- \end{matrix} \right)$ \\
\noalign{\smallskip}
 & $\delta \boldsymbol{y} = \text{const.}$ & $\delta \boldsymbol{y} = \text{const.}$ &  (observations at $\lambda \pm \delta \lambda$, $\delta \lambda / \lambda \ll 1$, $\partial_\lambda N_\mathrm{e}(\boldsymbol{x}_i,\lambda) \approx 0$) \\
 \noalign{\smallskip}
\hline
\end{tabular}
\tablefoot{For unknown sources, there is an additional degeneracy between the intrinsic source properties and the local lens properties given by ratios of second order derivatives of the deflection potential. Note the difference between $\boldsymbol{x}_i$ denoting image $i$ and $\boldsymbol{x}_{,i}$ denoting the $x_i$-coordinate of $\boldsymbol{x}$.}
\end{center}
\end{table*}

As stated in Table~\ref{tab:degeneracies}, the chance of observing the source in moving plasma lensing configurations offers the simplest way to break all degeneracies that occur in the lensing formalism. 
For plasma lensing configurations with unobserved source, we found the same degeneracies that also apply to gravitational lenses. 
But, due to the wavelength dependence of plasma lensing, additional ways to break these degeneracies become possible.
We therefore investigate the knowledge gain that can be obtained by having observations of a range of frequencies.

We start with the case in which no degeneracies occur to analyse the knowledge gain of the wavelength dependence. 
Having measured an observed arrival time difference $t$ for observed image and source positions $\boldsymbol{x}$ and $\boldsymbol{y}$, knowing the distances to the lens and the source, $\Gamma$ and a fixed ranges of wavelengths, Equation~\eqref{eq:t_plasma} allows us to locally probe $N_\mathrm{e}(\boldsymbol{x})$ over this wavelength range
\begin{equation}
N_\mathrm{e}(\boldsymbol{x},\lambda) =  \dfrac{2\pi}{r_\mathrm{e} \lambda^2} \left( \dfrac{c t}{1+z_\mathrm{d}} - \dfrac{D_\mathrm{d} D_\mathrm{s}}{D_\mathrm{ds}} \dfrac{(\boldsymbol{x}-\boldsymbol{y})^2}{2} \right)\;.
\label{eq:Ne_lambda}
\end{equation}
It thus reveals local absorption properties of the projected electron density at $\boldsymbol{x}$.
If the arrival time difference cannot be measured, e.g. because the source is not periodically time-varying, we can only probe the gradient with respect to $\boldsymbol{x}$ by means of Equation~\eqref{eq:alpha_plasma}
\begin{equation}
\nabla N_\mathrm{e}(\boldsymbol{x},\lambda) = -  \dfrac{2\pi}{r_\mathrm{e} \lambda^2} \dfrac{D_\mathrm{d} D_\mathrm{s}}{D_\mathrm{ds}} (\boldsymbol{x}-\boldsymbol{y}) \;.
\label{eq:gradNe_lambda}
\end{equation}
For $\lambda \rightarrow 0$, $\omega \rightarrow \infty$ in Equation~\eqref{eq:n_plasma}, there is a limiting frequency $\omega_t$ at which the transition from the strong to the weak lensing regime occurs, when we cannot resolve the multiple image positions or arrival time differences anymore.
Beyond $\omega_t$, $\boldsymbol{x} =  \boldsymbol{y}$, $t = 0$ and consequently, no multiple images can occur anymore, see also \cite{bib:Tsupko} for a recent illustration. 

The easiest way to break the degeneracy of Equation~\eqref{eq:los_degeneracy} is to observe plasma lenses in our cosmic neighbourhood, i.e.~for which $D_\mathrm{d} \rightarrow 0$, such that the observed arrival time difference is dominated by the plasma lens. 
The degeneracy is also broken for lenses very close to the source, $D_\mathrm{d} \rightarrow D_\mathrm{s}$, such that the geometric term dominates the observed arrival time difference.
\cite{bib:Brisken} and \cite{bib:Main} show example lenses with dominating geometric term.
Details about the ranges in which each term dominates are found in \cite{bib:Er4}.

For a general plasma lensing configuration, using Equation~\eqref{eq:Ne_lambda}, we now assume that we have three measurements of the left-hand sides at $\lambda_- \equiv \lambda - \delta \lambda$, $\lambda$, and $\lambda_+ \equiv \lambda + \delta \lambda$ with $\delta \lambda / \lambda \ll 1$. 
For each of these wavelengths, we assume that the arrival time difference and the image position relative to the source position are measured. 
Subtracting $N_\mathrm{e}(\boldsymbol{x},\lambda)$ from $N_\mathrm{e}(\boldsymbol{x},\lambda_+)$ and $N_\mathrm{e}(\boldsymbol{x},\lambda_-)$, we obtain
\begin{align}
N_\mathrm{e}(\boldsymbol{x},\lambda_+) - N_\mathrm{e}(\boldsymbol{x},\lambda) &\approx \delta \lambda \partial_\lambda N_\mathrm{e}(\boldsymbol{x},\lambda) \\
&= \dfrac{2\pi}{r_\mathrm{e} \lambda^2} \left( \dfrac{c \delta t_+}{1+z_\mathrm{d}} - \dfrac{D_\mathrm{d} D_\mathrm{s}}{D_\mathrm{ds}} \delta \boldsymbol{x}_+ (\boldsymbol{x}-\boldsymbol{y}) \right) \\
N_\mathrm{e}(\boldsymbol{x},\lambda_-) - N_\mathrm{e}(\boldsymbol{x},\lambda) &\approx -\delta \lambda \partial_\lambda N_\mathrm{e}(\boldsymbol{x},\lambda) \\
&= \dfrac{2\pi}{r_\mathrm{e} \lambda^2} \left( \dfrac{c \delta t_-}{1+z_\mathrm{d}} - \dfrac{D_\mathrm{d} D_\mathrm{s}}{D_\mathrm{ds}} \delta \boldsymbol{x}_- (\boldsymbol{x}-\boldsymbol{y}) \right) \;,
\end{align}
in which we denote $\delta t_{+/-} = t_{+/-}  - t$ and $\delta \boldsymbol{x}_{+/-} = \boldsymbol{x}_{+/-} - \boldsymbol{x}$.
Adding these equations, we can solve for the distance factor given the measured quantities, so that wavelength measurements are thus able to break the degeneracy of Equation~\eqref{eq:los_degeneracy}
\begin{equation}
\Gamma = \dfrac{\left(\delta t_+ + \delta t_- \right)}{\left(\delta \boldsymbol{x}_+ + \delta \boldsymbol{x}_- \right)\left(\boldsymbol{x}-\boldsymbol{y}\right)} \;.
\label{eq:break_Gamma}
\end{equation}
Since this calculation employs the assumption that the distances are neither dependent on $\boldsymbol{x}$, nor $\lambda$, a distance measured which is not based on luminosity (as the one mentioned in Section~\ref{sec:los_degeneracies}) should be used.
Without arrival time information, the degeneracy cannot be broken, as $\Gamma_\mathrm{p}$ is a common prefactor of all measurements in Equation~\eqref{eq:gradNe_lambda}.

Next, we want to break the degeneracy of Equation~\eqref{eq:constr_tau} by using multi-wavelength observations of $\tau$ and $(\boldsymbol{x}_1 - \boldsymbol{x}_2)$. 
We use measurements at three wavelengths as already introduced above with analogous notation and start from Equation~\eqref{eq:tau_plasma} solved for $N_-(\boldsymbol{x},\lambda) \equiv N_\mathrm{e}(\boldsymbol{x}_1) - N_\mathrm{e}(\boldsymbol{x}_2)$ 
\begin{align}
N_-(\boldsymbol{x},\lambda) =  \dfrac{2\pi}{r_\mathrm{e} \lambda^2} \dfrac{c \tau}{(1+z_\mathrm{d})} + \dfrac{\left( \boldsymbol{x}_1 - \boldsymbol{x}_2 \right)}{2} \nabla N_+(\boldsymbol{x},\lambda) \;,
\end{align}
in which $N_+(\boldsymbol{x},\lambda) \equiv N_\mathrm{e}(\boldsymbol{x}_1) + N_\mathrm{e}(\boldsymbol{x}_2)$.
Subtracting this Equation for the different wavelengths we arrive at 
\begin{align}
\delta \lambda \partial_\lambda A(\boldsymbol{x},\lambda) &=  \dfrac{2\pi}{r_\mathrm{e} \lambda^2} \dfrac{c \delta \tau_+}{(1+z_\mathrm{d})} + \dfrac{\delta \boldsymbol{x}_+}{2} \nabla N_+(\boldsymbol{x},\lambda) \label{eq:sys1} \\
-\delta \lambda \partial_\lambda A(\boldsymbol{x},\lambda) &=  \dfrac{2\pi}{r_\mathrm{e} \lambda^2} \dfrac{c \delta \tau_-}{(1+z_\mathrm{d})} + \dfrac{\delta \boldsymbol{x}_-}{2} \nabla N_+(\boldsymbol{x},\lambda) \;, \label{eq:sys2}
\end{align}
in which we have abbreviated 
\begin{equation}
A(\boldsymbol{x},\lambda) \equiv N_-(\boldsymbol{x},\lambda) - \dfrac{\left( \boldsymbol{x}_1 - \boldsymbol{x}_2 \right)}{2} \nabla N_+(\boldsymbol{x},\lambda)
\end{equation}
and assumed that the distance between the multiple images is altered by $\delta \boldsymbol{x}_{+/-}$ due to the change of wavelength.

Assuming that $\partial_\lambda A(\boldsymbol{x},\lambda) \approx 0$, i.e.~choosing $\delta \lambda$ such, that  $\partial_\lambda N_\mathrm{e}(\boldsymbol{x}) \approx 0$, we can use Equations~\eqref{eq:sys1} and \eqref{eq:sys2} to solve $\nabla N_+(\boldsymbol{x},\lambda)/2$ as the average gradient of the two image positions
\begin{equation}
\dfrac{1}{2} \nabla N_+(\boldsymbol{x},\lambda) = - \dfrac{2\pi}{r_\mathrm{e} \lambda^2} \dfrac{c}{(1+z_\mathrm{d})}  d^{-1}\left( \begin{matrix} \delta \boldsymbol{x}_{-,2} & -\delta \boldsymbol{x}_{+,2} \\ -\delta \boldsymbol{x}_{-,1} & \delta \boldsymbol{x}_{+,1}\end{matrix} \right) \left( \begin{matrix} \delta \tau_+ \\ \delta \tau_- \end{matrix} \right)
\label{eq:break_gradNe}
\end{equation}
with the determinant of the distance measurement matrix
\begin{equation}
d \equiv \left( \delta \boldsymbol{x}_{+,1} \delta \boldsymbol{x}_{-,2} -\delta \boldsymbol{x}_{+,2}\delta \boldsymbol{x}_{-,1} \right) \;,
\end{equation}
the subscripts denoting the vector components.
Fixing $\nabla N_+(\boldsymbol{x},\lambda)/2$ in this way, we break the degeneracy in Equation~\eqref{eq:constr_tau}.
Yet, the freedom remains to set one global constant in $N_\mathrm{e}(\boldsymbol{x})$, as discussed in Section~\ref{sec:overdensities}.

To summarise our findings, we add the degeneracy-breaking requirements to the right-most column of Table~\ref{tab:degeneracies}.

\section{Accuracy test by a simulated plasma lens}
\label{sec:example}

\begin{figure}[t!]
\centering
  \includegraphics[width=0.45\textwidth]{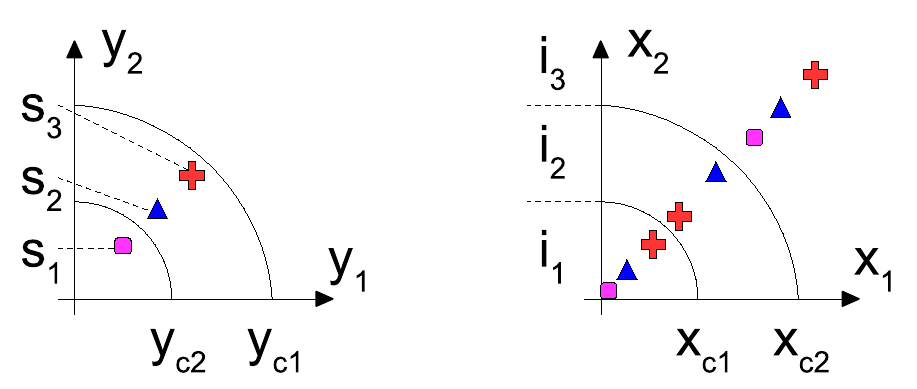} 
    \caption{Schematic distribution of source positions in the source plane (left axes) and respective image positions in the lens plane (right axes) for the Gaussian plasma lens discussed in Section~\ref{sec:example}. The source s$_1$ is placed at radial position $0.9 y_{c2}$, s$_2$ at $1.1 y_{c2}$, and s$_3$ at $0.9 y_{c1}$. For increasing observation frequency, the critical curves are shifted towards each other, decreasing the region for the images i$_2$.} 
 \label{fig:example_configs}
\end{figure}

In order to investigate the accuracy of the approximations made in Equations~\eqref{eq:break_Gamma} and \eqref{eq:break_gradNe}, we simulate an axisymmetric Gaussian plasma lens according to the estimates of \cite{bib:Clegg}
\begin{equation}
\psi_\mathrm{p} = \dfrac{D_{ds}}{D_d D_s} {r_\mathrm{e} \lambda^2 \over 2\pi} N_0 \; \exp \left(-\dfrac{\boldsymbol{x}^2}{2\sigma^2} \right) = \Gamma_\mathrm{p} N_\mathrm{e}(\boldsymbol{x})
\label{eq:Gauss_lens2}
\end{equation}
with $N_0=0.1~\text{pc}/\text{cm}^{-3}$ and $\sigma=1~ \text{AU}$ (astronomical unit). 
We furthermore assume that the plasma lens is located in our Galaxy at $D_\mathrm{d}=0.5~\text{kpc}$ (with $z_\mathrm{d} \approx 0$).
 
Using this lens, we investigate its impact in two plasma lensing configurations, one with a source located at distance $D_\mathrm{s} = 1~\text{kpc}$ in our Galaxy, and a second one with a source located at redshift $z_\mathrm{s} = 0.2$, which yields a typical extra-Galactic distance to a fast radio burst. 
For both sources, we determine the critical curves and caustics of the Gaussian lens and place the source relative to the caustics to obtain the three different image configurations summarised in Fig.~\ref{fig:example_configs}.
Source position~1 yields a highly demagnified central image which may be below the detection limit. 

To investigate the influence of the observation frequency, we assume two different main observation frequencies $\nu_1=0.5~\text{GHz}$ and $\nu_2=1.5~\text{GHz}$ with their corresponding wavelengths $\lambda_1$ and $\lambda_2$. 
For each wavelength, we employ $\delta \lambda$ approximately belonging to a frequency change of $\delta \nu_a \approx 5~\text{MHz}$, $\delta \nu_b \approx 10~\text{MHz}$, $\delta \nu_c \approx 20~\text{MHz}$, and $\delta \nu_d \approx 50~\text{MHz}$.
For each combination $\lambda_i \pm \delta \lambda_j$, $i=1,2$, $j=a,b,c,d$, we can calculate all source and image positions and arrival time differences to determine $\Gamma_\mathrm{sim}$ by means of Equation~\eqref{eq:break_Gamma} and compare it to the actual $\Gamma$ calculated from the distance ratio (see Equation~\eqref{eq:time_delay1}).
Analogously, we can determine the average electron density gradient from all combinations of multiple images for each source by means of Equation~\eqref{eq:break_gradNe}, here abbreviated as $\eta \equiv 1/2 \nabla N_+(\boldsymbol{x},\lambda)$ and compare $\eta_\mathrm{sim}$ to the actual average gradient $\eta$ as given by the derivatives of the Gaussian lens at the multiple image positions.
Due to the axisymmetry of this lens, Equation~\eqref{eq:break_gradNe} reduces to one equation, instead of two, $r$ being the radial position variable
\begin{equation}
\eta \equiv \dfrac12 \partial_r N_+(r,\lambda) \stackrel{!}{=} \dfrac{2\pi}{r_\mathrm{e}\lambda^2} \cdot \dfrac{c}{1+z_\mathrm{d}} \cdot \dfrac{\delta t_+ +  \delta t_-}{\delta r_+ + \delta r_-} \equiv \eta_\mathrm{sim}\;.
\label{eq:abbreviations_gradNe}
\end{equation}

\begin{figure*}[t!]
\centering
  \includegraphics[width=0.33\textwidth]{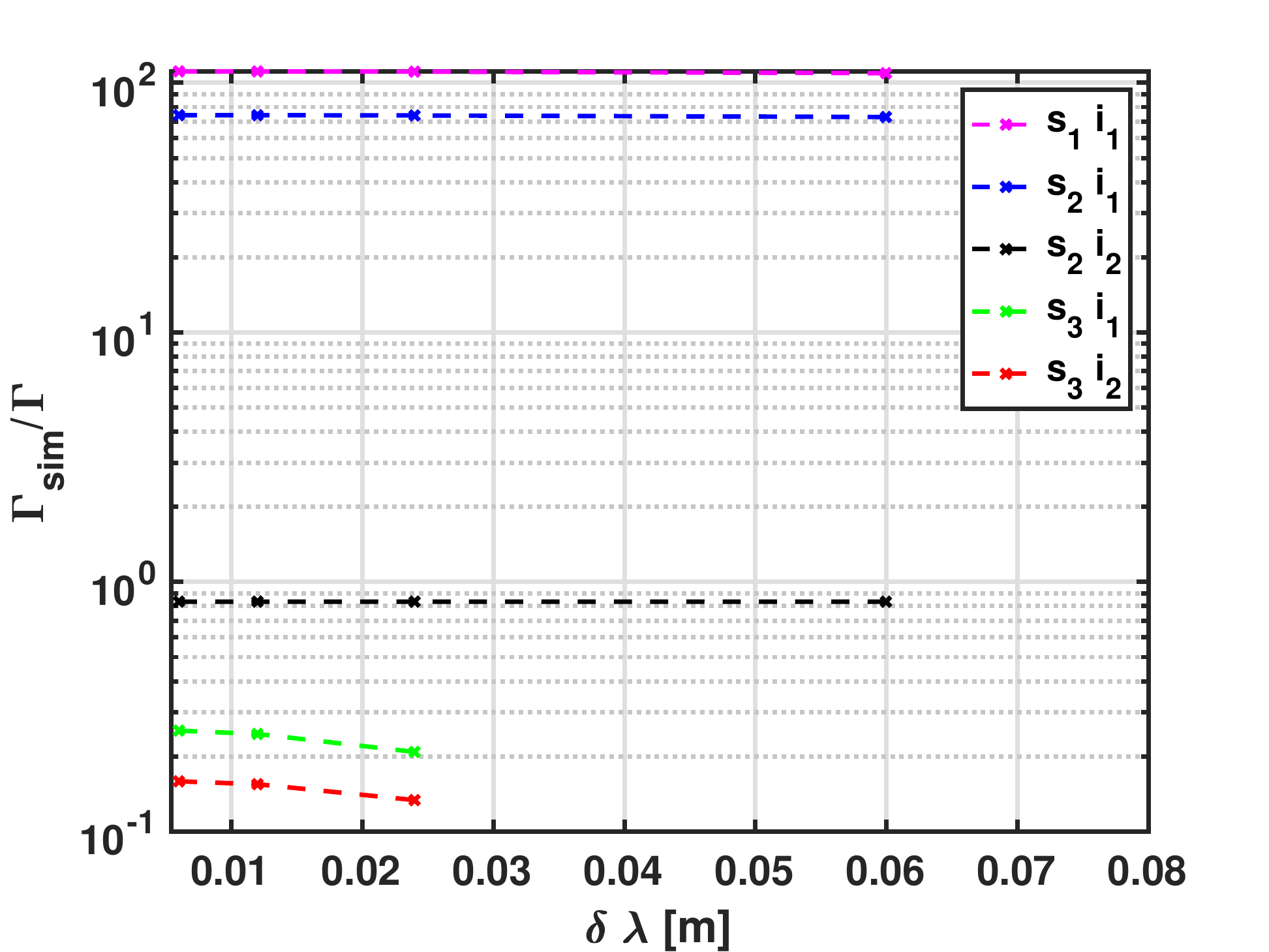} 
  \includegraphics[width=0.33\textwidth]{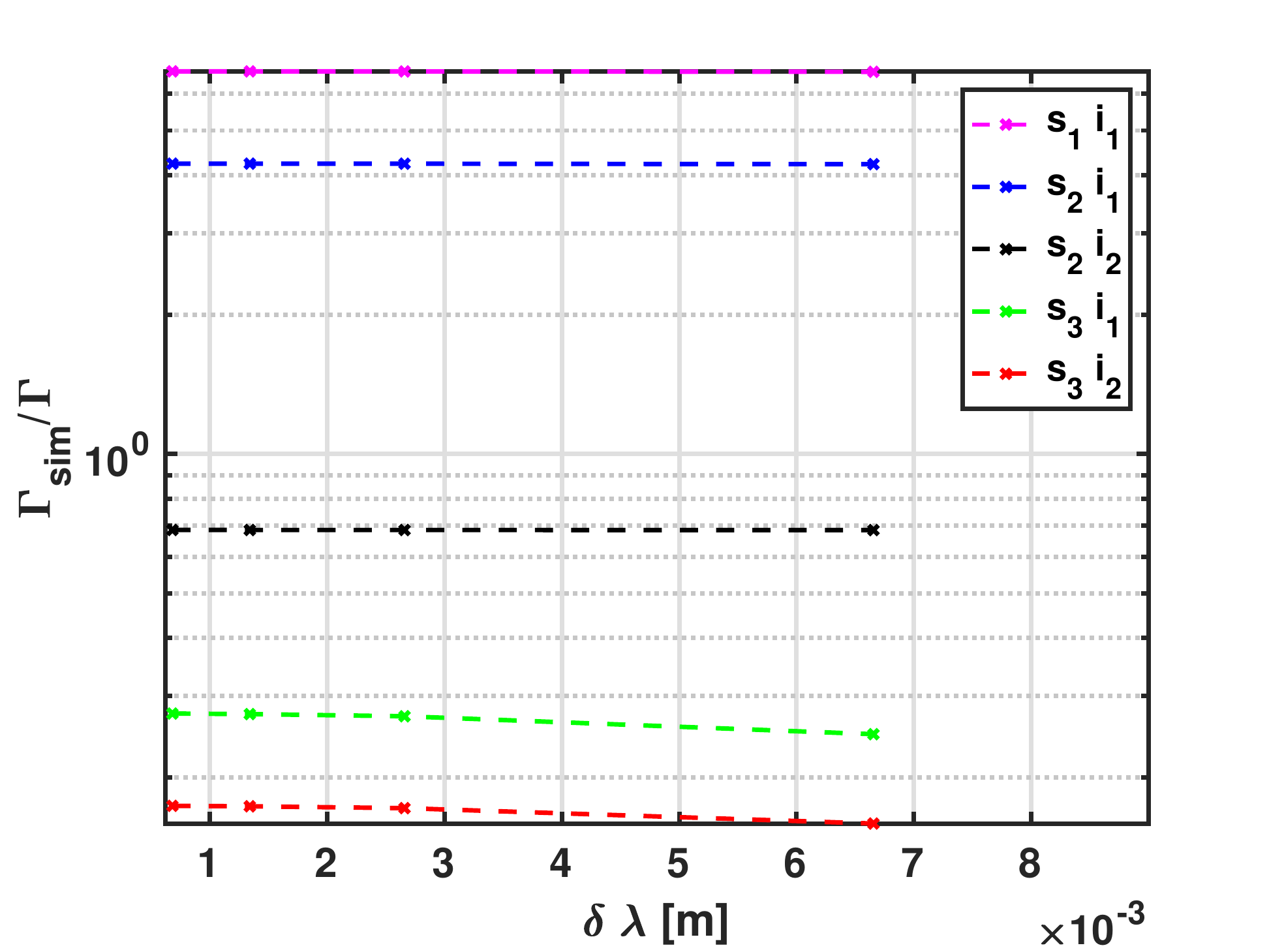} 
  \includegraphics[width=0.33\textwidth]{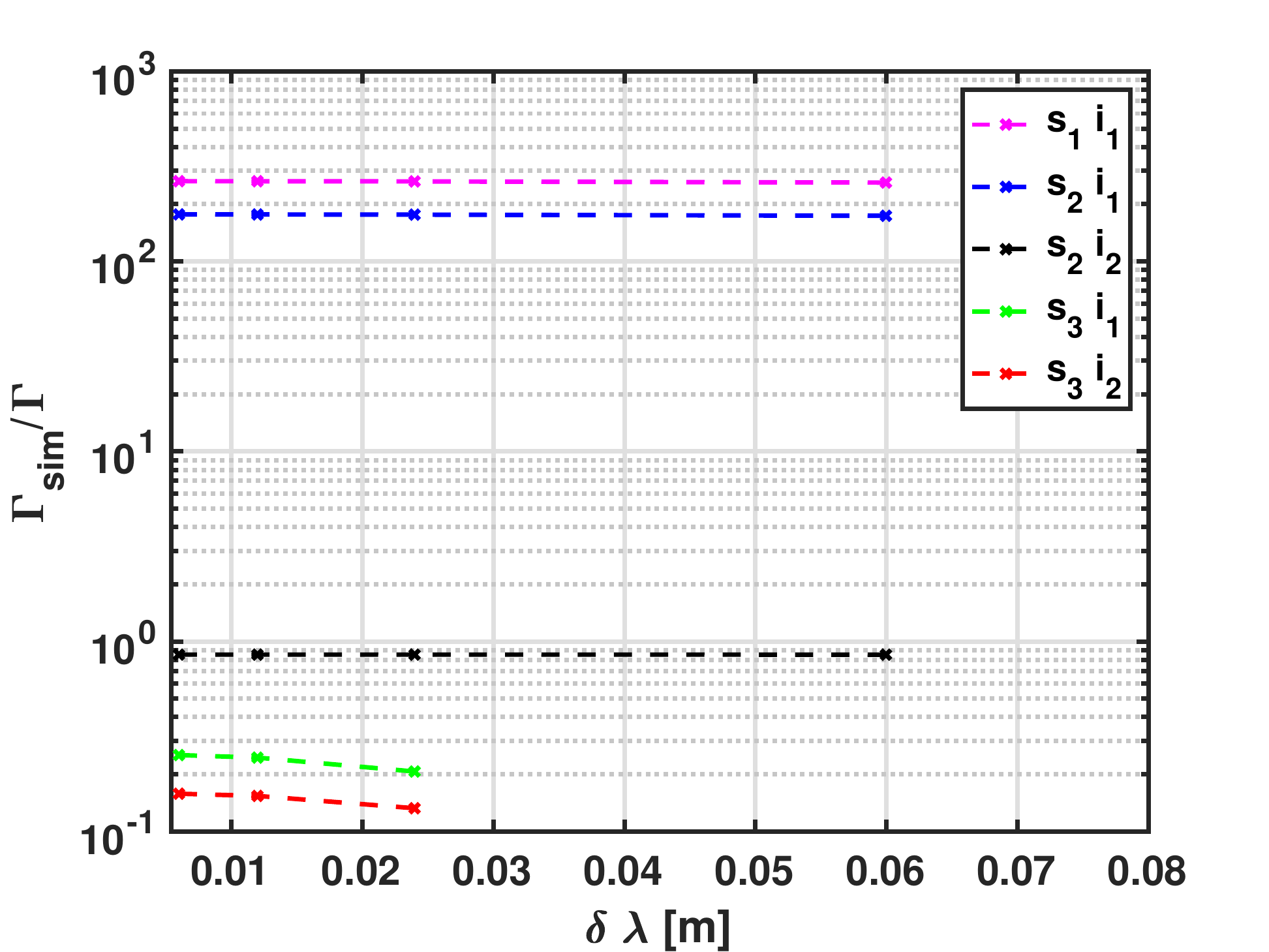} 
    \caption{Accuracy test of Equation~\eqref{eq:break_Gamma}: ratio of the simulated $\Gamma_\mathrm{sim}$ to the true $\Gamma$ (defined in Equation~\eqref{eq:time_delay1}) for the lens detailed in Equation~\eqref{eq:Gauss_lens2}, $D_\mathrm{s}=1~\text{kpc}$ and $\nu_1=0.5~\text{GHz}$ (left), $\nu_2=1.5~\text{GHz}$ (centre), and for a source at $z_\mathrm{s}=0.2$ and $\nu_1=0.5~\text{GHz}$ (right). The different lines belong to the images $i_j$, $j=1,2,3$ for the three different source positions $s_1$, $s_2$, and $s_3$ as shown in Figure~\ref{fig:example_configs}.} 
 \label{fig:results_gamma}
\end{figure*}

Figure~\ref{fig:results_gamma} summarises the accuracy of Equation~\eqref{eq:break_Gamma} for the Galactic source for $\nu_1$ (left), $\nu_2$ (centre) and for the fast radio burst as a source for $\nu_1$ (right). 
For $\nu_1$, the image positions for $\delta \nu_d$ could not be robustly determined and are therefore excluded from the analysis. 
Yet, we expect the curves to further decrease.
Comparing the left and central plot for $\nu_1$ and $\nu_2$, we find that the overall accuracy to approximate $\Gamma$ by $\Gamma_\mathrm{sim}$ increases. 
The highest accuracy in reconstruction is observed for image~2 of source~2, which is the image closest to the inner boundary of the outer critical curve (see Figure~\ref{fig:example_configs}). For increasing proximity to this critical curve, which occurs in our setting for decreasing $\nu$, the accuracy is increased.
As the comparison of the left and the right plot shows, the increasing distance of the fast radio burst leads to an increase in the reconstruction accuracy of $\Gamma_\mathrm{sim}$ for image~2 of source~2, but worse reconstruction accuracies for the other images.
The constraints on the required precision of the observables to arrive that this accuracy is summarised in Table~\ref{tab:break_results} together with $\Gamma_\mathrm{sim}/\Gamma$ for the most accurate image configuration. As noted in Section~\ref{sec:observables_plasma}, the astrometric measurements are the most challenging ones.  
 
\begin{figure*}[h!]
\centering
  \includegraphics[width=0.33\textwidth]{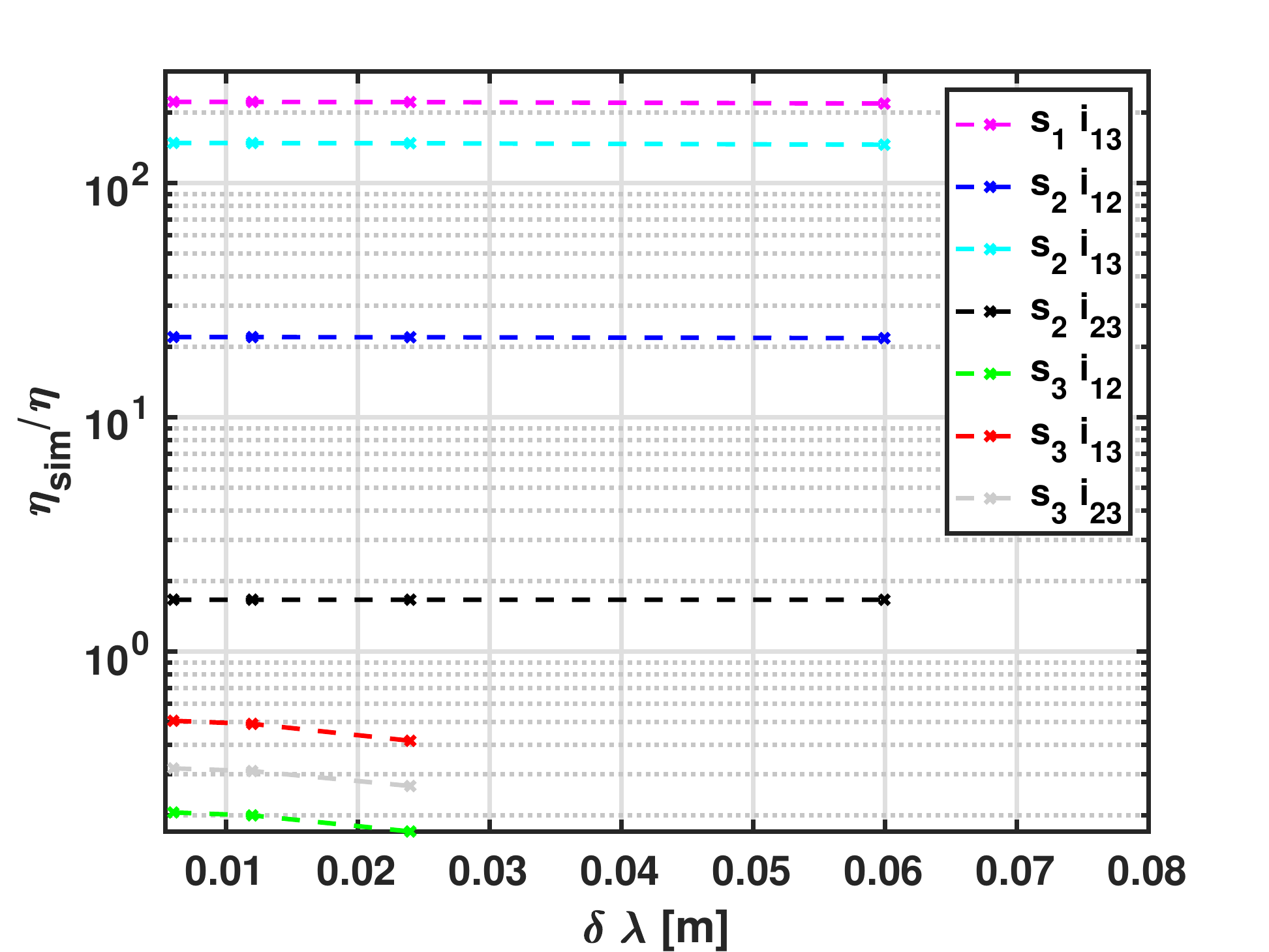} 
  \includegraphics[width=0.33\textwidth]{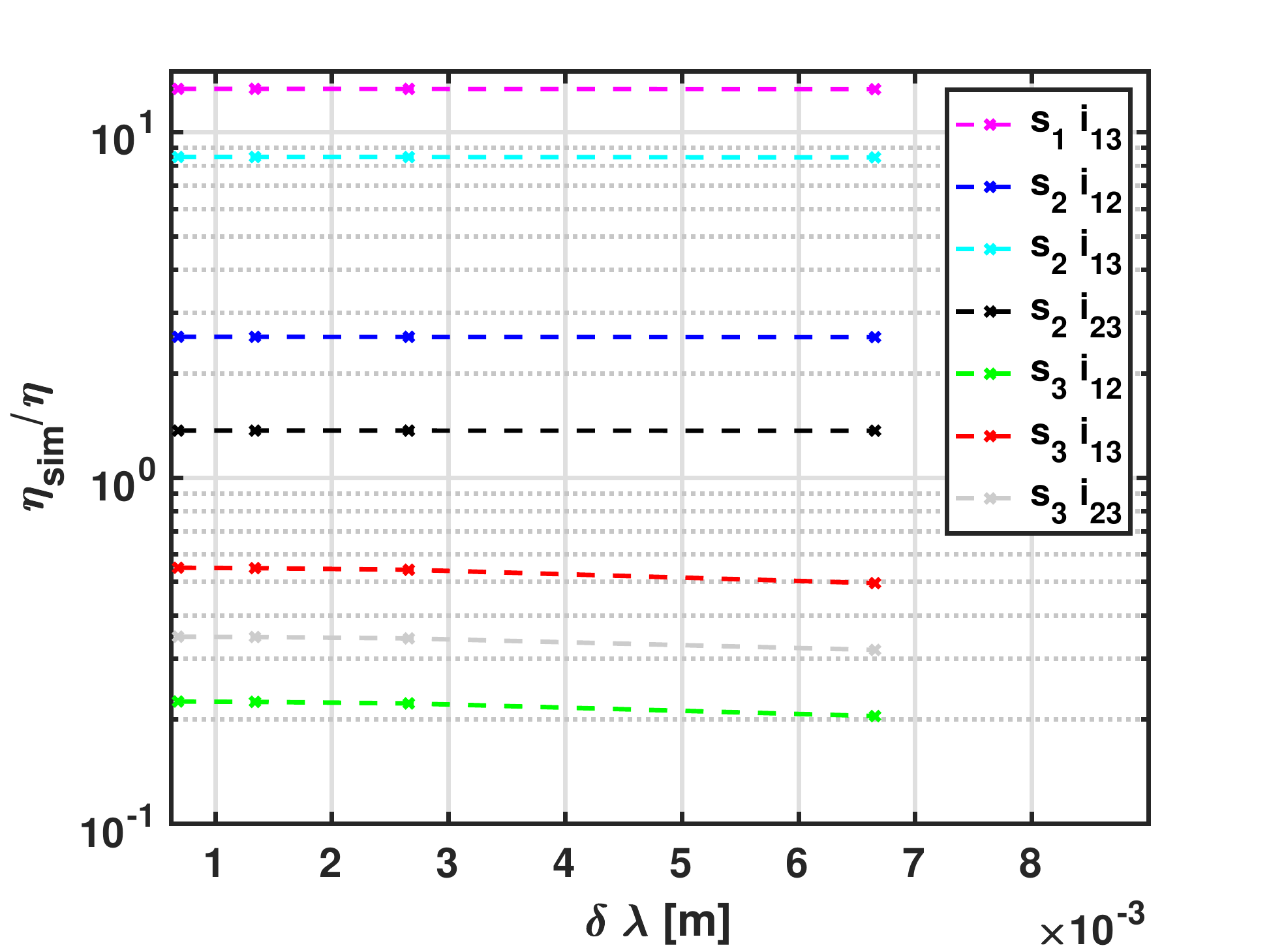} 
  \includegraphics[width=0.33\textwidth]{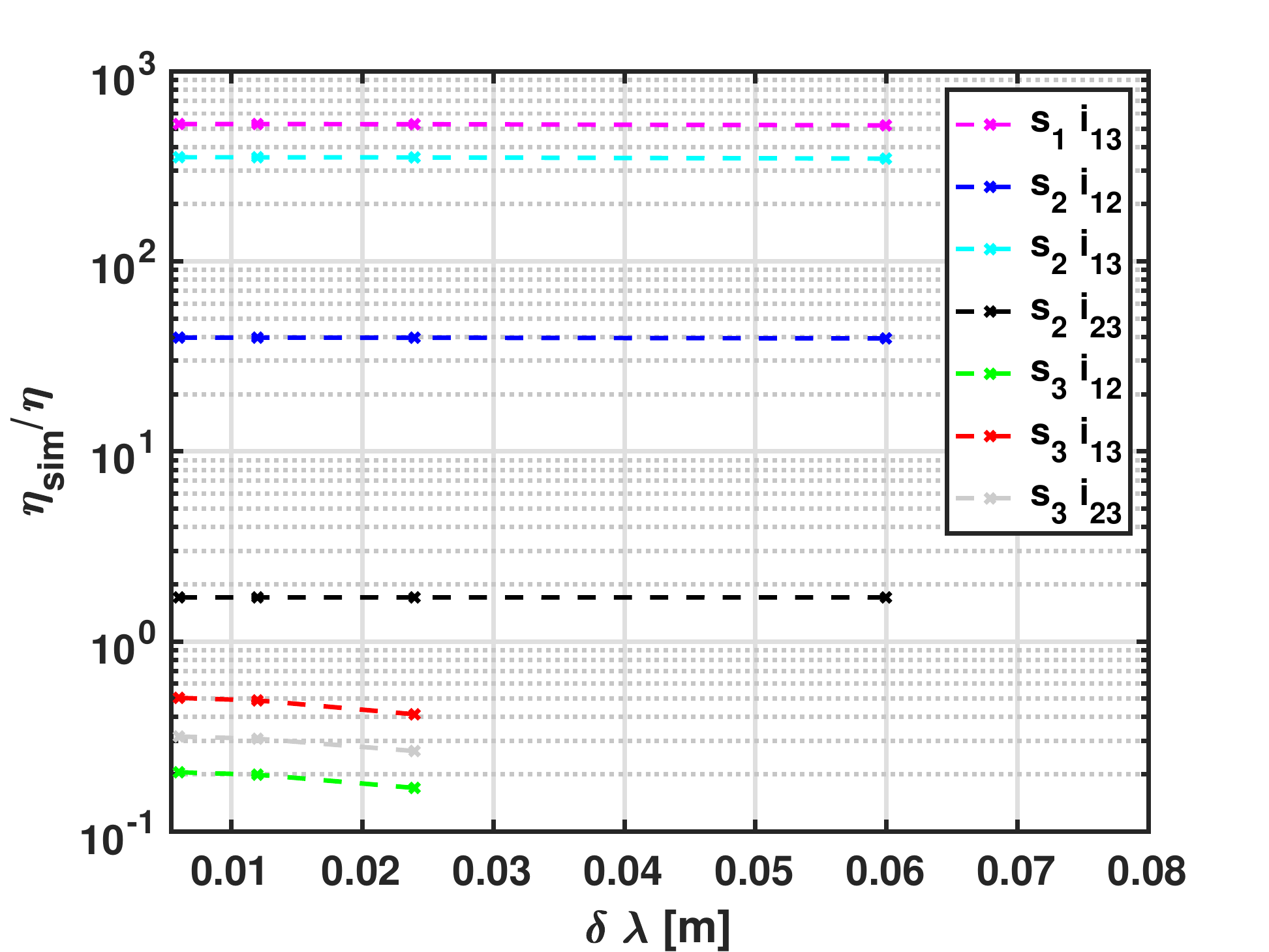}  
    \caption{Accuracy test of Equation~\eqref{eq:break_gradNe} with $\eta_\mathrm{sim}$ and $\eta$ defined in Equation~\eqref{eq:abbreviations_gradNe}, same configurations as in Figure~\ref{fig:results_gamma}. The different lines $i_{jk}$, $j,k=1,2,3$, belong to the pairs of multiple images $i_j$ and $i_k$ used to determine the arrival time difference and distance between the images.} 
 \label{fig:results_gradNe}
\end{figure*}

In Figure~\ref{fig:results_gradNe} the accuracy of Equation~\eqref{eq:break_gradNe} is plotted for the Galactic source for $\nu_1$ (left), $\nu_2$ (centre) and for the fast radio burst as a source for $\nu_1$ (right). 
We systematically test the accuracy for all possible multiple image pairs $i_j$, $i_k$, $j,k=1,2,3$, abbreviated as $i_{jk}$, for each source position. 
As in Figure~\ref{fig:results_gamma}, we find that the overall reconstruction accuracy is increased for increasing observation frequency $\nu$ when comparing the left and the central plot. 
The combination of images~2 and 3 for source~2 yields the highest accuracy for the mean electron density gradient.  
The accuracy increases for decreasing distance between $i_2$ and $i_3$, which occurs for increasing observation frequency $\nu$. 
Comparing the left and the right plot, we find that the increasing source distance slightly deteriorates the accuracy at the same observation frequency. 
Table~\ref{tab:break_results} also lists the requirements on the observation precision to arrive at the accuracies in Figure~\ref{fig:results_gradNe} and shows the ratios $\eta_\mathrm{sim}/\eta$ for the most accurate lens configuration.
The arrival time difference $\tau_{23}$ is equal to the arrival time difference between image~2 and the source, $t_2$, because image~3 is at the source position and the lens is weak, such that the arrival time differences are dominated by the geometric part. 

\begin{table}[t]
 \caption{Accuracy of Equations~\eqref{eq:break_Gamma} and \eqref{eq:break_gradNe} for source~2 of Figure~\ref{fig:example_configs}, values of observable properties to determine $\Gamma_\mathrm{sim}$ and $\eta_\mathrm{sim}$, and radii of the critical curves of the Gaussian plasma lens.}
\label{tab:break_results}
\begin{center}
\begin{tabular}{r|ccc}
\hline
\noalign{\smallskip}
	&	$\nu=\nu_1$	&	$\nu=\nu_2$	&	$\nu=\nu_1$	\\
Property & $D_\mathrm{s}=1~\text{kpc}$ & $D_\mathrm{s}=1~\text{kpc}$ & $z_\mathrm{s}=0.2$ \\ 
\hline
$\Gamma_\mathrm{sim}/\Gamma(\delta \nu_d, i_2)$	&	0.83	& 0.68	&	0.85	\\
$\eta_\mathrm{sim}/\eta(\delta \nu_d,i_{23})$	&	1.67	&	1.37	&	1.70	\\
$x_2/x_{c2}$	&	0.98	&	0.95	&	0.98	\\
$t_2$ [$\mu$s]	&	345	&	38	&	342	\\
$\tau_{23}$ [$\mu$s]	&	345	&	38	&	342	\\
$|x_2-y|$ [mas]	&	-17.17	&	-6.36	&	-24.13	\\
$x_2(\lambda_d)-x_2(\lambda)$ [mas]	&	0.15	&	0.08	&	0.14	\\
$x_{c1}$ [mas]	&	1.95	&	1.99	&	1.95	\\
$x_{c2}$ [mas]	&	5.62	&	4.64	&	5.89	\\
\hline
\end{tabular}
\end{center}
\end{table}

\begin{table*}[t]
 \caption{Synopsis and comparison of gravitational and plasma lensing as a theory of light deflection by a deflection potential in the limit of geometrical optics.}
\label{tab:comparison}
\begin{center}
\begin{tabular}{rcc}
\hline
\noalign{\smallskip}
\textbf{Property} & \textbf{Gravitational lensing} & \textbf{Plasma lensing} \\
\noalign{\smallskip}
\hline
\noalign{\smallskip}
Cause of deflection ($\boldsymbol{r} \in \mathbb{R}^3$) & inhomogeneous mass density $\rho(\boldsymbol{r})$ & inhomogeneous charge density $n_\mathrm{e}(\boldsymbol{r})$\\
\noalign{\smallskip}
Projected quantity ($\boldsymbol{x} \in \mathbb{R}^2$) & surface mass density $\Sigma(\boldsymbol{x})$ & surface charge number density $N_\mathrm{e}(\boldsymbol{x})$ \\
\noalign{\smallskip}
\hline
\noalign{\smallskip}
Relation to projected & Poisson equation & Proportionality \\
deflection potential $\psi(\boldsymbol{x})$ &  $\Delta \psi_\mathrm{m}(\boldsymbol{x}) = 2 \tfrac{\Sigma(\boldsymbol{x})}{\Sigma_\mathrm{c}}$ & $\psi_\mathrm{p}(\boldsymbol{x}) = - \tfrac{D_\mathrm{ds}}{D_\mathrm{s} D_\mathrm{d}} \tfrac{\lambda^2 r_\mathrm{e}}{2\pi} N_\mathrm{e}(\boldsymbol{x})$ \\
\noalign{\smallskip}
\hline
\noalign{\smallskip}
Type of lens w.r.t.~vacuum & converging: $\psi_\mathrm{m}(\boldsymbol{x}) > 0$ & diverging: $\psi_\mathrm{p}(\boldsymbol{x}) < 0$ \\
Arrival time difference & retarded: $t_\mathrm{p} - t_\mathrm{u} \propto \Sigma(\boldsymbol{x})$ & advanced: $t_\mathrm{p} - t_\mathrm{u} \propto -|N_\mathrm{e}(\boldsymbol{x})|$ \\
\noalign{\smallskip}
\hline
\noalign{\smallskip}
Wavelength dependence of $\psi(\boldsymbol{x})$ & achromatic: independent & chromatic: $\psi(\boldsymbol{x}) \propto \lambda^2$ \\ 
\noalign{\smallskip}
\hline
\noalign{\smallskip}
Polarisation change  & negligible for static lenses & Faraday rotation by a magnetic field \\
\noalign{\smallskip}
\hline
\noalign{\smallskip}
Scale of appearance & critical surface mass density  & plasma frequency \\
 &  $\Sigma_\mathrm{c} = \tfrac{c^2}{4\pi G}\tfrac{D_\mathrm{s}}{D_\mathrm{d} D_\mathrm{ds}}$ & $\omega_p \equiv \sqrt{4\pi e^2 n_e/m_e}$  \\
\noalign{\smallskip}
\hline
\noalign{\smallskip}
Degeneracies & unbroken due to & broken by \\
 & unobservable source & multi-wavelength observations \\
\noalign{\smallskip}
\hline
\end{tabular}
\end{center}
\end{table*}

\section{Conclusion}
\label{sec:conclusion}

In this work, we unified the geometric optics formalism to describe light deflection by a single thin static deflection screen for gravitational and plasma lensing. 
The light propagation for both cases is assumed to come from a source at potentially extra-galactic but finite distance.
The light rays travel through the deflecting screen and, after passing it, follow a deflected path until observed.
Based on this common formalism, we showed that gravitational and plasma lensing are subject to the same degeneracies, but, that plasma lensing degeneracies can be broken more easily. The quantitative methodology to break the plasma lensing degeneracies can be found in Table~\ref{tab:degeneracies} (for a table of all definitions, see Table~\ref{tab:abb}) which lead to the following conclusions:

Plasma lensing configurations may be transient, such that the source becomes observable. 
For galaxy-scale and galaxy-cluster-scale gravitational lensing configurations, this scenario is impossible during human life-times. 
Directly observing the source breaks all occurring degeneracies, if the background cosmology is known and provides distances to the lens and the source. 

Plasma lensing is wavelength-dependent, such that we can determine the distance ratio given by Equation~\eqref{eq:break_Gamma} for the case of an observable source. 
For the case of an unobservable source, we can determine the average gradient in the plasma electron density by Equation~\eqref{eq:break_gradNe}. 
Both results are obtained based on the observational evidence of arrival time differences and relative image positions without employing a specific model for the deflecting electron density distribution. 
The highest accuracy of the inferred distance ratio and the average gradient in the plasma electron density is obtained for the multiple image closest to the inner boundary of the outer critical curve. 
Since the accuracy is dependent on the main observing frequency, we can actively increase it by selecting the frequency accordingly. 
Section~\ref{sec:example} shows an example case of a Gaussian lens, in which we find that the astrometric measurements have to reach milli-arcsecond precision or better to obtain accuracies of about 80\% in the distance ratio and more than 60\% in the average plasma density gradient.
For gravitational lenses, these degeneracies can only be broken by additional assumptions like a deflecting mass density model and a cosmological model to set up a distance measure or by using complementary observations.

In addition, we showed that a Gaussian plasma lens with vanishing width results in a point plasma, see Section~\ref{sec:point_plasma} for details. 
Consequently, assembling a sophisticated plasma lens as a Gaussian mixture model finds its analogue in assembling a gravitational lens by individual point masses. Table~\ref{tab:comparison} summarises and compares the properties of gravitational and plasma lensing as further detailed in Section~\ref{sec:differences}.

With this unified formalism and a systematic analysis of its degeneracies specified to plasma lenses of geometric optics, we can now start analysing observational cases like the one discussed in \cite{bib:Brisken}.
In addition, the approach can be extended to account for the relative motion of the lens and source and the corresponding flux density changes over time, i.e.~we can compare plasma lensing to the transient phenomenon of gravitational micro-lensing. 
As observational evidence suggests, see e.g.~\cite{bib:Liu} or \cite{bib:Simard2}, the approach may also be extended to thick lenses or multiple lens plane, which has already been studied in gravitational lensing as well and found to be highly degenerate \cite{bib:Wagner4}, \cite{bib:Schneider}.

Since some plasma lenses seem to show image configurations caused by a combination of diffraction and refraction like discussed in \cite{bib:Pen2} or \cite{bib:Grillo}, we can also investigate the information gain when including wave-optic phenomena into our approach. 
Additional polarisation measurements can be included as well to probe the magnetic field in the plasma (\cite{bib:Li}).

We are thus very confident that joining all these observations at the positions of observed images will yield enough information to set up a data-driven, global model of the deflecting plasma. Hence, although gravitational lensing is always praised by its simplicity not including any complicated baryonic physics in its formalism, it is exactly this multitude of baryonic interactions which allows us to break the formalism-intrinsic degeneracies in a simple and observation-based way.

\begin{acknowledgements}
JW gratefully acknowledges the support by the Deutsche Forschungsgemeinschaft (DFG) WA3547/1-3, the hospitality of the SWIFAR during her research stay in which most of the ideas were worked out, and thanks Jori Liesenborgs for helpful discussions.
XE is supported by NSFC grant No. 11873006.
\end{acknowledgements}

\bibliographystyle{aa}
\bibliography{plasma}

\appendix
\section{Abbreviations and definitions}
\label{app:abb}

Table~\ref{tab:abb} lists all definitions of abbreviations that we introduced and employed.

\begin{table}[h!]
 \caption{Table of definitions of abbreviations that we employ.}
\label{tab:abb}
\begin{center}
\begin{tabular}{r|cc}
\hline
\noalign{\smallskip}
\begin{tabular}{r} Description \end{tabular} & Name & Definition\\ 
\hline
\begin{tabular}{r} Distance ratio \end{tabular} & $\Gamma$ & $\dfrac{(1+z_\mathrm{d})}{c} \dfrac{D_\mathrm{d} D_\mathrm{s}}{D_\mathrm{ds}}$ 
\\[2ex]
\begin{tabular}{r} 
Prefactor of \\ plasma potential \end{tabular} & $\Gamma_\mathrm{p}$ &  $\dfrac{D_\mathrm{ds}}{D_\mathrm{d} D_\mathrm{s}} \dfrac{r_\mathrm{e} \lambda^2}{2\pi}$ 
\\[2ex]
\begin{tabular}{r} Prefactor of $\tau$ \\ for plasma lenses \end{tabular} & $\Gamma_\tau = \Gamma \, \Gamma_\mathrm{p}$ &  $\dfrac{(1+z_\mathrm{d})}{c} \dfrac{r_\mathrm{e} \lambda^2}{2\pi}$ \\ 
\hline
\begin{tabular}{r} Difference betw. transf. \\ and original variable $v$ \end{tabular} &  $\delta v$ &  $\tilde{v}- v$
\\[2ex]
\begin{tabular}{r} $i$th coordinate \\ of a vector $\boldsymbol{v}$ \end{tabular} & $\boldsymbol{v}_{,i}$ & 
\\
\hline
\begin{tabular}{r} Wavelength variable \end{tabular} & $\lambda_\pm$ & $\lambda \pm \delta \lambda$
\\[1ex]
\begin{tabular}{r} Wavelength-dependent \\ difference of a variable $t$ \end{tabular} & $\delta t_\pm$ & $t(\lambda \pm \delta \lambda) - t(\lambda)$  
\\[2ex]
\begin{tabular}{r} Electron density variable \end{tabular} & $N_\pm(\boldsymbol{x},\lambda)$ & $N_\mathrm{e}(\boldsymbol{x}_1) \pm N_\mathrm{e}(\boldsymbol{x}_2)$ 
\\[1ex]
\begin{tabular}{r} Observable electron \\ density gradient \end{tabular} & $\eta$ & $\tfrac12 \nabla N_+(\boldsymbol{x},\lambda)$
\\
\hline
\end{tabular}
\end{center}
\end{table}

\section{Change of the reference value for the refractive index}
\label{app:shifting_n}

Not choosing the vacuum as a reference to determine the refractive index, Equation~\eqref{eq:t_psi} reads
\begin{equation}
\hat{t}_\psi = \int \limits_{|\boldsymbol{d}_\mathrm{OS}|}^{0} \mathrm{d} l  \, \left( \dfrac{1}{v(\boldsymbol{r})} - \dfrac{1}{v_\mathrm{ref}} \right) \;,
\end{equation}
in which $v(\boldsymbol{r})$ denotes the phase velocity of the light ray propagating through the deflecting structure and $v_\mathrm{ref}$ the constant phase velocity of the reference light ray, propagating through a homogeneous medium with constant refractive index.
For vacuum, $v_\mathrm{ref} = c$.
The reference light ray is only necessary as a basis for comparison to determine the arrival time difference with respect to it, so that the observed quantity is $\hat{t}_\psi = \hat{t}_\mathrm{p} - \hat{t}_\mathrm{u}$.

Adding and subtracting the travel time of an unperturbed path in vacuum, we obtain
\begin{align}
\hat{t}_\psi &= \int \limits_{|\boldsymbol{d}_\mathrm{OS}|}^{0} \mathrm{d} l  \, \left(  \left( \dfrac{1}{v(\boldsymbol{r})} - \dfrac{1}{c} \right) - \left( \dfrac{1}{v_\mathrm{ref}} - \dfrac{1}{c} \right) \right) \\
&= \dfrac{(1+z_\mathrm{d})}{c} \left( \tilde{\psi}(\boldsymbol{x}) - \tilde{\psi}_\mathrm{ref} \right) \label{eq:ref} \;,
\end{align}
such that the arrival time difference as compared to Equation~\eqref{eq:t_psi} is changed by a constant offset, represented by the last term in Equation~\eqref{eq:ref}. 
This term can be interpreted as the difference in travel time of a light ray in the homogeneous medium with constant refractive index compared to a light ray in vacuum. 
Inserting $\hat{t}_\psi$ as given by Equation~\eqref{eq:ref} into Equation~\eqref{eq:time_delay2} or \eqref{eq:time_delay3} to obtain an arrival time difference between two multiple images, the constant offset cancels out.
Analogously, Equation~\eqref{eq:lens_equation1} and all equations derived from it remain unaffected by the constant offset, as they only contain derivatives of the deflection potential.
For Equation~\eqref{eq:time_delay1}, the change of the unperturbed path of the reference light ray, does not affect the physical conclusions, as long as the geometric light deflection is consistently taking place on the same background. 

\end{document}